\documentclass[aps,prepr,twocolumn,showpacs,superscriptaddress,floatfix]{revtex4-2}
\usepackage[T1]{fontenc}
\usepackage[utf8]{inputenc}
\usepackage[normalem]{ulem}
\usepackage{amsmath,amssymb, bm}

\usepackage{graphicx}
\graphicspath{{Figures/}}
\usepackage{dcolumn}

\usepackage[dvipsnames]{xcolor}
\usepackage[colorlinks=true,
  citecolor=Cerulean,
  linkcolor=RubineRed,
  urlcolor=Cerulean]{hyperref}

\usepackage{comment}

\begin{document}
\title{Floquet engineering of spin-valley selective transport in  jacutingaite}
	
\author{Otman Bouladiane}
\author{Kamal Azaidaoui}
\affiliation{Laboratory of Theoretical Physics, Faculty of Sciences, Choua\"ib Doukkali University, PO Box 20, 24000 El Jadida, Morocco}
\author{Clarence Cortes}
\affiliation{Vicerrector\'ia de Investigaci\'on y Postgrado, Universidad de La Serena, La Serena 1700000, Chile}  
\author{David Laroze}
\affiliation{Instituto de Alta Investigación, Universidad de Tarapacá, Casilla 7D, Arica, Chile}

\author{Ahmed Jellal}
\email{a.jellal@ucd.ac.ma}
\affiliation{Laboratory of Theoretical Physics, Faculty of Sciences, Choua\"ib Doukkali University, PO Box 20, 24000 El Jadida, Morocco}

\begin{abstract}
We study electron transport through a monolayer jacutingaite (Pt$_2$HgSe$_3$) tunnel junction in which only the barrier is irradiated by off-resonant circularly polarized light, while the leads remain undriven. In the high-frequency regime, the driven barrier reduces to an effective static Dirac Hamiltonian with a photon-dressed, valley-dependent mass term. A staggered sublattice potential $V_z$ and a substrate-induced exchange field $m_s$ provide additional tunable mass terms. Using scattering theory, we compute spin- and valley-resolved transmission and reflection, as well as the Landauer conductance. Photon dressing shifts the barrier {Dirac masses} with opposite signs in the ($\bm K,\bm K'$) valleys and induces a splitting of the propagation thresholds. The finite barrier then produces channel-dependent Fabry--P\'erot-type interference through the phase $q_x^{\eta s_z}L$. We find broad parameter windows with near-perfect valley filtering ($|P_v|\simeq 100\%$) and substantial spin polarization ($|P_s|\sim 70\%$). The dominant spin and valley polarizations can be switched by tuning the drive amplitude $A_0$, $V_z$, and~$m_s$.
\end{abstract}

\maketitle

\section{Introduction}
\label{sec:introduction}


Topological insulators represent an intriguing material class because their edge states withstand nonmagnetic disordering effects~\cite{Hasan2010,Qi2011}. The quantum spin Hall phase demonstrates this behavior through its two-dimensional edge states, which exhibit helical {propagation} that connects their spin and momentum. The binding of these two properties prevents electrons from experiencing backward scattering because time-reversal symmetry {forbids such processes}~\cite{Kane2005a,Kane2005b,Bernevig2006}.
The existing protection for this system requires that materials need to generate bulk gaps of substantial size because this feature enables them to function properly at higher temperatures. The performance of graphene technology shows its most significant limitation at this point. Graphene exists as a gapless semimetal with Dirac fermions, but its intrinsic spin-orbit coupling capabilities remain extremely weak, which leads to an almost nonexistent gap~\cite{CastroNeto2009,Yao2007}. Silicene provides better performance than other materials because its buckled structure improves spin-orbit coupling, although the resulting gap from this improvement stays very small at only several meV~\cite{Molle2017,Ezawa2012}.
Three methods can be used to improve gap control: external field application, engineered structure development, and material coupling. These methods offer two key advantages: they secure topological phases and enable enhanced control over spin-dependent transport, a vital requirement for future electronic and spintronic technologies.


{
Experimentally, the quantum spin Hall (QSH) effect has been confirmed to date on a small set of platforms, each trading the gap size against the robustness of the material. The effect was first observed in inverted HgTe/CdTe quantum wells~\cite{Konig2007} and subsequently in InAs/GaSb bilayers~\cite{Knez2011}. However, the inverted gaps in these heterostructures are only a few meV to a few tens of meV, restricting helical edge transport to sub-kelvin or few-kelvin temperatures. Monolayer WTe$_2$ extended helical conduction to temperatures of order 100~K, a substantial improvement, but the material is highly air-sensitive, particularly at the monolayer limit, complicating device fabrication and long-term operation \cite{Fei2017,Wu2018,Ye2016}. Bismuthene grown epitaxially on SiC opened a much larger gap of about 0.8~eV, enabling QSH signatures at room temperature, but it exists only as a substrate-bound epitaxial layer and cannot be exfoliated or freely transferred \cite{Reis2017}. These platforms illustrate a persistent trade-off between gap size, environmental stability, and material processability. Jacutingaite is attractive precisely because it promises to relax this trade-off: its predicted gap of about 0.5~eV is comparable to that of bismuthene, yet, unlike bismuthene, it is a naturally occurring van der Waals crystal that has been reported to be stable under ambient conditions and can be mechanically exfoliated into thin flakes, with STM and ARPES measurements providing experimental evidence for a large-gap electronic structure consistent with the QSH phase \cite{Vymazalova2012,Kandrai2020,Cucchi2020}. This combination of a sizable, robust gap with standard 2D-material processability is what motivates using jacutingaite as a platform for the tunable, Floquet-engineered spin-valley transport studied in this work.
}

Jacutingaite (Pt$_2$HgSe$_3$) is a promising alternative material~\cite {Vymazalova2012}. It has been predicted to realize a Kane-Mele quantum spin Hall insulator with a sizable band gap of order $0.5$~eV~\cite{Marrazzo2018}. This prediction has motivated rapid experimental work. Scanning tunneling microscopy (STM) experiments reported signatures consistent with a large-gap quantum spin Hall phase~\cite{Kandrai2020}, and angle-resolved photoemission spectroscopy (ARPES) subsequently mapped the electronic structure of Pt$_2$HgSe$_3$~\cite{Cucchi2020}. Jacutingaite forms a layered mineral with a buckled honeycomb lattice~\cite{Vymazalova2012}. It remains stable under ambient conditions and can be exfoliated into thin flakes~\cite{Longuinhos2020}. {Exfoliated crystals down to one or two layers have also been reported to remain stable under ambient exposure for months~\cite{Kandrai2020}}. Additionally, it has been proposed to host dual topological phases, making it an attractive platform for tunable topological devices~\cite{Facio2019}.
Another factor that makes jacutingaite noteworthy is its tunability. A perpendicular electric field induces a staggered sublattice potential, which breaks inversion symmetry. As a result, the system can be tuned between quantum spin Hall, quantum valley Hall, and trivial insulating phases~\cite{Marrazzo2018, Rehman2022, Vargiamidis2022}. Exchange fields provide another route for control. They lift spin degeneracy and shift the phase boundaries~\cite{Vargiamidis2022}. Jacutingaite therefore offers a clean setting to study how competing mass terms reshape the low-energy Dirac bands.

Circularly polarized light provides dynamic control beyond static gating. In the off-resonant regime, the driven system can be described by an effective time-independent Hamiltonian. {The light appears as an additional valley-dependent mass term that breaks time-reversal symmetry.} This mechanism {underlies} Floquet topological phases in Dirac materials~\cite{Oka2009, PhysRevB.84.235108, Ezawa2013, Rudner2020}. {Driven} jacutingaite has been studied in several settings~\cite{tsrz-5t2s,skyn-fprn}. The interplay of irradiation with additional symmetry-breaking fields produces a rich phase structure. {It also leads to} regimes with strong spin and valley selectivity~\cite{Alipourzadeh2023, shah2024topological, Hajati2025}. {Device operation, however, is ultimately} a transport question. {It requires} currents and conductances, not only bulk invariants.
On the transport side, Floquet engineering has advanced considerably. In the high-frequency regime, a periodically driven system can remain in a long-lived prethermal state.  One can then proceed with methods that are familiar from equilibrium transport~\cite{Bukov2015}. In particular, scattering theory provides a direct route to calculating conductances and channel-resolved transmission. Floquet scattering makes this connection explicit for driven systems~\cite{Li2018}. It also allows one to formulate topological information directly in terms of a scattering matrix~\cite{Fulga2016}. Time-dependent barrier geometries are especially useful in this context{, 
 they generate sidebands and enable}  photon-assisted tunneling, and can yield strong spin and valley selectivity~\cite{Jongchotinon2020}. Much of the existing work, however, assumes homogeneous driving or considers idealized protocols.

The device-relevant setting is often different. In many junctions, only part of the sample is exposed to light, resulting in a spatially inhomogeneous Floquet problem. One then has undriven leads connected to a driven barrier region. In the barrier, the Dirac masses become photon-dressed. In the leads, the asymptotic states remain those of the equilibrium material. Similar junctions in other two-dimensional systems have demonstrated valley-selective and spin-valley-selective filtering by combining gating with irradiation~\cite{Liu2021, Hajati2021}. 
In addition, asymmetric tunnel junctions have been shown to exhibit pronounced magnetoresistance effects~\cite{Qiu2020}, and proximity-induced exchange fields offer an extra knob, enabling large spin-valley polarizations in transport~\cite{Niu2019, Hajati2025}. 
These findings strongly motivate a systematic exploration of analogous junction setups in jacutingaite, where the interplay of light, gating, and exchange fields could be exploited for tunable topological transport.


Motivated by the potential for controllable spin and valley currents in two-dimensional topological materials, as well as the unique tunability offered by photon-dressed Dirac masses in jacutingaite, we study spin-valley selective transport in monolayer jacutingaite tunnel junctions. We consider a simple planar setup in which an undriven source and drain sandwich a finite barrier region that is irradiated by off-resonant circularly polarized light, while the leads remain in equilibrium. In addition, the barrier hosts a staggered sublattice potential and a substrate-induced exchange field. 
In the high-frequency limit, the driven barrier can be mapped onto an effective static system. This yields spin- and valley-dependent Dirac masses, which directly influence the transport properties. 
{We stress that this bulk-transport approach probes the same Kane--Mele spin--orbit coupling responsible for the QSH phase. However, it probes this coupling through bulk band splitting and mass engineering rather than through protected edge conduction. The relation between the two is made explicit in Sec.\ref{sec:theory}.}
Using scattering theory, we calculate the transmission, reflection, and conductance across the junction. This allows us to uncover regimes of pronounced spin and valley polarization. We also demonstrate how these can be tuned through the driving amplitude and barrier parameters. Thus, this setup offers a practical platform to investigate dynamically controlled, highly selective transport with clear implications for future spintronic and topological device applications.

{We stress that a photon-induced valley mass is not unique to jacutingaite. Studies of circularly polarized light have shown that it also lifts the valley degeneracy in irradiated $\alpha$-$T_3$ lattices~\cite{PhysRevB.98.075422,PhysRevB.99.205135}. In these systems, the low-energy theory is a three-band pseudospin-one Dirac--Weyl Hamiltonian: it carries a valley
index, while the electron spin enters only as a degeneracy factor. Such a drive cannot split the two spin species, so the resulting transmission and conductance selectivity is intrinsically valley-only. In jacutingaite, by
contrast, the intrinsic spin--orbit coupling has the Kane--Mele form $\eta s_z\lambda_{\rm so}$, which directly links the valley index $\eta$ to the physical electron spin $s_z$. As shown explicitly in Sec.~\ref{sec:theory}\,B, this allows the drive to produce genuine spin polarization even without an exchange field -- an effect forbidden in a spin-degenerate model -- and
gives rise to a four-channel (spin$\times$valley) selectivity mechanism with no counterpart in the two-channel, valley-only selectivity reported for $\alpha$-$T_3$ junctions.
}


The remainder of the paper is organized as follows. In Sec.~\ref{sec:theory}, we present a theoretical model and describe the scattering theory for the junction.  Sec.~\ref{sec:Tran_COndu} focuses on the spin- and valley-resolved transmission and reflection, highlighting regimes with strong channel selectivity. 
In Sec.~\ref{CCC}, we calculate and discuss the corresponding conductance. While
Sec.~\ref{sec:Polarization} examines the spin and valley polarization of the transport. Finally, Sec.~\ref{sec:conclusion} summarizes our results and outlines their potential implications in device applications.

\section{Model and methods}
\label{sec:theory}

\subsection{Junction geometry}

\begin{figure}[ht!]
    \centering
    \includegraphics[width=\columnwidth]{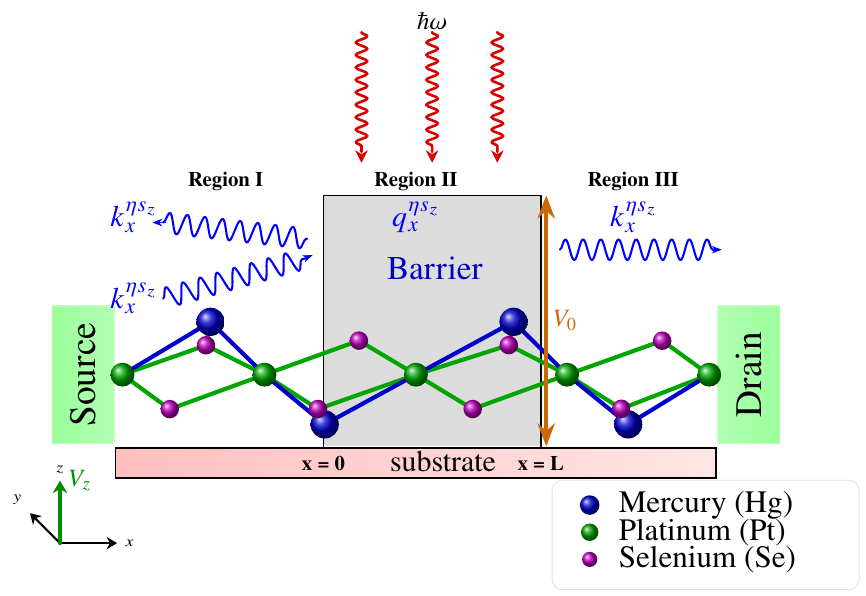}
    \caption{Schematic of the tunnel junction. Region~II, the barrier of width $L$, is irradiated by circularly polarized light ($\hbar\omega$), and includes a scalar potential $V_0$, a staggered sublattice potential $V_z$, and a spin-dependent exchange field $m_s$. Regions~I and III are undriven source and drain leads.
}
    \label{fig:schema}
\end{figure}

We consider a planar tunnel junction in monolayer jacutingaite. The system is translationally invariant along {the} $y$-direction {and} piecewise uniform along the transport $x$-direction. 
The junction consists of three regions, as depicted in Fig.~\ref{fig:schema}. Region I ($x<0$) is the source lead{, region} III ($x>L$) is the drain lead{, and both are undriven}. Region II ($0\le x\le L$) is the active barrier {of width} $L${, with a} scalar potential of height $V_0${, and is} irradiated by off-resonant circularly polarized light. In the off-resonant regime, the drive generates an effective valley-dependent term in the low-energy theory. We also include a staggered sublattice potential $V_z${, which breaks} inversion symmetry and provides an additional control knob. Finally, we include a substrate-induced exchange term that couples to spin and sublattice, parametrized by $m_s$.

\subsection{Effective Floquet Hamiltonian} 

We consider the low-energy Hamiltonian around the $\bm K$ and $\bm K'$ valleys \cite{Hajati2025,tsrz-5t2s,skyn-fprn,shah2024topological}. We index the valleys by $\eta=\pm 1$, with $\bm K$ ($+1$) and $\bm K'$ ($-1$), and spin by $s_z=\pm 1$, with $+1$ for spin-up and $-1$ for spin down. The Pauli matrices $\sigma_i$$(i=x,y,z)$ will be used to describe the sublattice degree of freedom, while $\sigma_0$ denotes the $2 \times 2$ identity matrix. To include the effect of an external drive, we introduce a time-dependent field, which makes the Hamiltonian periodic in time. This provides a simple way to capture how the drive influences the electronic structure and the resulting topological features of the system. 
Let us start from the undriven Hamiltonian
\begin{align}
H^{\eta s_z}_{0}
=\hbar v_F (\eta k_x\sigma_x+k_y\sigma_y)+\Delta^{\eta s_z}\sigma_z + V(x)\sigma_0,
\label{eq:ham0}
\end{align}
where $k_x$ and $k_y$ are momenta measured from the Dirac points, and  the Dirac mass takes the form
\begin{align}
\Delta^{\eta s_z}=\eta s_z \lambda_{\mathrm{so}} + s_z m_{s} + V_z.
\label{eq:gap}
\end{align}
Here, $\lambda_{\mathrm{so}}=81.2~\mathrm{meV}$ is the intrinsic spin-orbit coupling \cite{shah2024topological,Hajati2025,4jx2-11ks}. The term $s_z m_s$ models a substrate-induced exchange field. The parameter $V_z$ describes a staggered sublattice potential induced by a perpendicular electric field $E_z$ \cite{Hajati2025,shah2024topological}. We use $v_F=3.0\times 10^{5}~\mathrm{ms^{-1}}$ \cite{shah2024topological,4jx2-11ks}.
We take the scalar barrier potential $V(x)$  as
\begin{align}
V(x)=
\begin{cases}
0, & \text{I, } x<0,\\
V_0, & \text{II, } 0\le x\le L,\\
0,\  & \text{III, } x>L.
\end{cases}
\label{eq:profile}
\end{align}

We couple the drive through minimal substitution, $\bm k\to \bm k+\frac{e}{\hbar}\bm A(t)$. 
We take a circularly polarized vector potential in the $x$-$y$ plane \cite{PhysRevB.103.245432,Qiu2020,PhysRevB.84.235108},
\begin{align}
\bm A(t)=A_0\bigl(\gamma \sin\omega t,\ \cos\omega t,\ 0\bigr),
\label{eq:A}
\end{align}
where $\gamma=\pm 1$ denotes the helicity, with $\gamma=+1$ for right circular polarization and $\gamma=-1$ otherwise. The amplitude is $|A_0|=|E_0|/\omega$, with $E_0$ the electric-field amplitude. Now the Hamiltonian takes the form
\begin{widetext}
\begin{align}
H^{\eta s_z}=
\hbar v_F\Big[\eta\Big(k_x+\frac{e}{\hbar^2}A_x(t)\Big)\sigma_x
+\Big(k_y+\frac{e}{\hbar^2}A_y(t)\Big)\sigma_y\Big]
+\Delta^{\eta s_z}\sigma_z+ V(x)\sigma_0.
\label{eq:hamt}
\end{align}
\end{widetext}
In the off-resonant, high-frequency regime, we describe the driven region by an effective static Floquet Hamiltonian. To leading order in $1/\omega$, the van Vleck (high-frequency) expansion yields \cite{PhysRevB.103.245432,PhysRevB.84.235108}
\begin{align}
H^{\eta s_z}_{\mathrm{eff}}
=H^{\eta s_z}_{0}(x)+\frac{[H_{-1},H_{1}]}{\hbar\omega}
+\mathcal{O}(\omega^{-2}),
\label{eq:Heff_expand}
\end{align}
where the Fourier components are
\begin{align}
H_{n}=\frac{\omega}{2\pi}\int_{0}^{\frac{2\pi}{\omega}}dt H^{\eta s_z}(t,x)e^{-in\omega t}.
\label{eq:Hn}
\end{align}
For the present model, the leading correction simplifies to a valley-dependent term that renormalizes the Dirac mass. We write \cite{PhysRevB.103.245432,Qiu2020}
\begin{align}
\frac{[H_{-1},H_{1}]}{\hbar\omega}=\eta \frac{\gamma \left(e A_{0} v_{F}\right)^{2}}{\hbar \omega}\sigma_z,
\label{eq:masslight}
\end{align}
We set $\lambda_{\omega}
=\frac{\gamma \left(e A_{0} v_{F}\right)^{2}}{\hbar \omega}$, which is
 a drive-induced effective term in the off-resonant regime. It is controlled by $(A_0,\omega)$ through the high-frequency expansion. {For the amplitudes used below, $\lambda_\omega\simeq0.6\,\lambda_{\mathrm{so}}$ at $A_0=0.4$~Vfs/nm and $\lambda_\omega\simeq2.3\,\lambda_{\mathrm{so}}$ at $A_0=0.8$~Vfs/nm, so the drive is comparable to the intrinsic spin--orbit scale in the first case and larger than it in the second. Eq.~\eqref{eq:effgap} also separates the roles of the three fields inside the barrier. Taking differences of the barrier masses gives $\Delta^{\bm K s_z}_{\mathrm{eff}}-\Delta^{\bm K' s_z}_{\mathrm{eff}}=2(\lambda_\omega+s_z\lambda_{\mathrm{so}})$ and $\Delta^{\eta\uparrow}_{\mathrm{eff}}-\Delta^{\eta\downarrow}_{\mathrm{eff}}=2(m_s+\eta\lambda_{\mathrm{so}})$, while $V_z$ enters all four with the same coefficient. The drive therefore sets the light-induced contribution to the valley splitting, the exchange field sets the explicit spin splitting, and the staggered potential shifts the whole set together.}
The resulting effective Hamiltonian in each region reads
\begin{align}
H^{\eta s_z}_{\mathrm{eff}}
=\hbar v_F(\eta k_x\sigma_x+k_y\sigma_y)
+\Delta^{\eta s_z}_{\mathrm{eff}}(x)\sigma_z
+V(x)\sigma_0,
\label{eq:ham3}
\end{align}
where we have defined  an effective Dirac mass as
\begin{align}
\Delta^{\eta s_z}_{\mathrm{eff}}(x)= \Delta^{\eta s_z}+\eta\lambda_\omega(x).
\label{eq:effgap}
\end{align}
%
These relations reveal the coupled roles of light, spin, and valley. Even without an exchange field ($m_s=0$), the mass splitting $\Delta_{\rm eff}^{\eta\uparrow}-\Delta_{\rm eff}^{\eta\downarrow}=2\eta\lambda_{\rm so}$ is already nonzero. The spin-dependent masses in jacutingaite are intrinsically split by the Kane--Mele spin--orbit coupling. The drive further reshapes this splitting through $\lambda_\omega$ and can enhance or suppress it depending on
its sign and amplitude. Because light and spin--orbit coupling add in one spin sector and compete in the other, the valley contrast can be confined to a single spin channel purely by tuning the drive amplitude $A_0$. This mechanism underlies the simultaneous, largely independent control of spin polarization $P_s$ and valley polarization $P_v$ demonstrated in Secs.~\ref{CCC} and~\ref{sec:Polarization}. It is most clearly visible in the operating maps of Figs.~\ref{fig:TVzms}, \ref{fig:GVzms}, and~\ref{fig:PVzms}. Such four-channel control has no counterpart in spin-degenerate, valley-selective Dirac models such as $\alpha$-$T_3$.
Throughout our study, we take $\hbar\omega=0.31~\mathrm{eV}$ as a representative off-resonant photon energy, corresponding to $\lambda\simeq4~\mu\mathrm{m}$ in the mid-infrared range used in strong-field TMD pump experiments~\cite{Kobayashi2023FloquetExcitonsWS2}.

The choice to irradiate only the barrier is deliberate. It generates the physics exploited throughout this work. Because the drive is confined to region~II, the lead mass remains at its undriven value $\Delta^{\eta s_z}$. In the barrier, the mass becomes $\Delta^{\eta s_z}_{\rm eff} = \Delta^{\eta s_z} + \eta\lambda_\omega(x)$. Thus, each interface has a finite mass mismatch of $\eta\lambda_\omega$, with opposite signs in the two valleys. This  mismatch is the key ingredient behind the valley- and spin-selective scattering discussed in Secs.~\ref{sec:Tran_COndu}--\ref{sec:Polarization}. Under uniform illumination of the entire junction, the same photon-dressed shift would affect both the leads and the barrier. The mass discontinuity would then vanish. Consequently, the interface-scattering mechanism responsible for the selectivity reported here would also disappear. The light would instead act as a global renormalization of the bulk mass rather than as a locally tunable control parameter \cite{AZAIDAOUI2026116579}. Locally illuminated junctions of this type are not purely theoretical. Comparable geometries, in which only part of a two-dimensional sample is driven while the rest remains in equilibrium, have been used to realize light-controlled and valley-selective transport in other systems \cite{PhysRevB.103.245432,Li2025SciRep,Liu2021}. 
On the experimental side, achieving a sharp, spatially confined drive is increasingly feasible. Mid-infrared fields can be confined to deep sub-wavelength regions in van der Waals materials through near-field and polaritonic focusing techniques \cite{Basov2016,Chen2012}, offering a concrete route to realizing the locally irradiated barrier geometry assumed in this work.

It is worth emphasizing how this bulk scattering problem connects to the QSH character of jacutingaite. In a ribbon or flake geometry, the Kane--Mele term $\eta s_z\lambda_{\rm so}$ in Eq.~\eqref{eq:gap} opens the bulk gap and supports helical, backscattering-protected edge states. These edge states are not considered here. Our junction is unbounded along $y$ and operates at energies above the bulk gap. The current is therefore carried by bulk propagating Dirac states rather than edge modes. The QSH physics nevertheless remains essential. The same spin--orbit term that opens the topological gap also introduces the valley-dependent contribution to $\Delta^{\eta s_z}$ in Eq.~\eqref{eq:gap}. It therefore makes the two valleys inequivalent in the undriven leads. Photon dressing and the exchange field then modify this intrinsic mass splitting. The drive can either reinforce or compete with $\lambda_{\rm so}$ in the two spin sectors, as shown by the mass differences following Eq.~(\ref{eq:effgap}). Jacutingaite is therefore more than a convenient Dirac material for this study. Its large Kane--Mele-derived spin--orbit scale, $\lambda_{\rm so}\simeq81$~meV, sets the scale of the bulk spin-valley selectivity reported in Secs.~\ref{sec:Tran_COndu}--\ref{sec:Polarization}. This occurs even though the transport considered here is mediated by bulk states rather than protected edge modes.

The closing conditions $\Delta^{\eta s_z}_{\mathrm{eff}}=0$ are more than propagation thresholds. Written out, they read
\begin{align}
V_z+s_z m_s=-\eta\left(s_z\lambda_{\mathrm{so}}+\lambda_{\omega}\right).
\label{eq:closing}
\end{align}
These are four straight lines in the $(V_z,m_s)$ plane, whose slope is fixed by the spin index and whose offset carries the valley index. Crossing one of them inverts the sign of a channel mass, and in jacutingaite such an inversion marks a topological phase boundary~\cite{Vargiamidis2022, Alipourzadeh2023,tsrz-5t2s}. In our junction, the same four lines organize the transmission maps shown in Fig.~\ref{fig:TVzms}. A channel sits deepest inside its propagating window near its own closing line, and turns evanescent once its barrier mass exceeds the threshold set by Eq.~\eqref{eq:qx}.

\subsection{Solutions of energy spectrum}

The junction is invariant along $y$, and therefore $k_y$ is a conserved quantity, i.e., $[H^{\eta s_z}_{\mathrm{eff}},k_y]=0$. Consequently, the spinor associated with the Hamiltonian Eq.~\eqref{eq:ham3} is separable in coordinates
$
\Psi^{\eta s_z}(x,y)=\phi^{\eta s_z}(x)e^{ik_y y}.$
From the eigenvalue equation
$H^{\eta s_z}_{\mathrm{eff}}\Psi^{\eta s_z}=E^{\eta s_z}\Psi^{\eta s_z}$, we show that 
in each region the spinors $\phi^{\eta s_z}(x)$ can be expressed as follows
\begin{align}
\phi_\text{I}^{\eta s_z}(x)
&=\chi_{1-}^{\eta s_z} e^{i k_{x}^{\eta s_z} x}
+ r^{\eta s_z}\chi_{1+}^{\eta s_z} e^{-i k_{x}^{\eta s_z} x},
\label{eq:psiI}\\
\phi_\text{II}^{\eta s_z}(x)
&=a^{\eta s_z}\chi_{2-}^{\eta s_z} e^{i q_{x}^{\eta s_z} x}
+ b^{\eta s_z}\chi_{2+}^{\eta s_z} e^{-i q_{x}^{\eta s_z} x},
\label{eq:psiII}\\
\phi_\text{III}^{\eta s_z}(x)
&=t^{\eta s_z}\chi_{1-}^{\eta s_z} e^{i k_{x}^{\eta s_z} x}.
\label{eq:psiIII}
\end{align}
Here $r^{\eta s_z}$ and $t^{\eta s_z}$ are the reflection and transmission amplitudes, and $a^{\eta s_z},b^{\eta s_z}$ are barrier-mode amplitudes.
We have  set
\begin{align}
\chi_{1,\pm}^{\eta s_z}=
\begin{pmatrix}
\mp \alpha^{\eta s_z}e^{\pm i\phi^{\eta s_z}}\\
1
\end{pmatrix},
\quad
\chi_{2,\pm}^{\eta s_z}=
\begin{pmatrix}
\mp \beta^{\eta s_z}e^{\pm i\theta^{\eta s_z}}\\
1
\end{pmatrix},
\label{eq:spinors}
\end{align}
where the incident $\phi^{\eta s_z}$ and inter-barrier $\theta^{\eta s_z}$  angles are given by
\begin{align}
&e^{\pm i\phi^{\eta s_z}}=\frac{\eta k_x^{\eta s_z}\pm ik_y}{\sqrt{(k_x^{\eta s_z})^2+k_y^2}},
\\
&e^{\pm i\theta^{\eta s_z}}=\frac{\eta q_x^{\eta s_z}\pm ik_y}{\sqrt{(q_x^{\eta s_z})^2+k_y^2}}.
\label{eq:angles}
\end{align}
The parameters $\alpha^{\eta s_z}$ and $\beta^{\eta s_z}$ encode the mass-dependent spinor structure. They are
\begin{align}
\alpha^{\eta s_z}
&=
\frac{
\Delta^{\eta s_z}
+\sqrt{(\Delta^{\eta s_z})^{2}+(\hbar v_F)^2\big[(k_x^{\eta s_z})^2+k_y^2\big]}
}{
\hbar v_F\sqrt{(k_x^{\eta s_z})^2+k_y^2}
},
\label{eq:alpha}\\
\beta^{\eta s_z}
&=
\frac{
\Delta^{\eta s_z}_{\mathrm{eff}}
+\sqrt{(\Delta^{\eta s_z}_{\mathrm{eff}})^{2}+(\hbar v_F)^2\big[(q_x^{\eta s_z})^2+k_y^2\big]}
}{
\hbar v_F\sqrt{(q_x^{\eta s_z})^2+k_y^2}
}.
\label{eq:beta}
\end{align}
The longitudinal wave vectors in the leads and the barrier follow from the dispersion. In regions I and III,  we obtain
\begin{align}
k_x^{\eta s_z}
=\frac{1}{\hbar v_F}\sqrt{(E^{\eta s_z})^{2}-(\Delta^{\eta s_z})^{2}-(\hbar v_F k_y)^2},
\label{eq:kx}
\end{align}
while in region II, we have
\begin{align}
q_x^{\eta s_z}
=\frac{1}{\hbar v_F}\sqrt{(E^{\eta s_z}-V_0)^{2}-(\Delta^{\eta s_z}_{\mathrm{eff}})^{2}-(\hbar v_F k_y)^2}.
\label{eq:qx}
\end{align}
By considering the barrier profile $V(x)$~\eqref{eq:profile}, we write the corresponding energy as
\begin{align}
E^{\eta s_z}_s = V(x)+ s\sqrt{(\hbar v_F)^2\big[(k_x^{\eta s_z})^2+k_y^2\big]+\big(\Delta^{\eta s_z}_{\mathrm{eff}}(x)\big)^2},
\label{eq:disper}
\end{align}
where $s=\pm$ labels conduction ($+$) and valence ($-$) bands.

\begin{figure*}[ht!]
\centering
\includegraphics[scale=0.44]{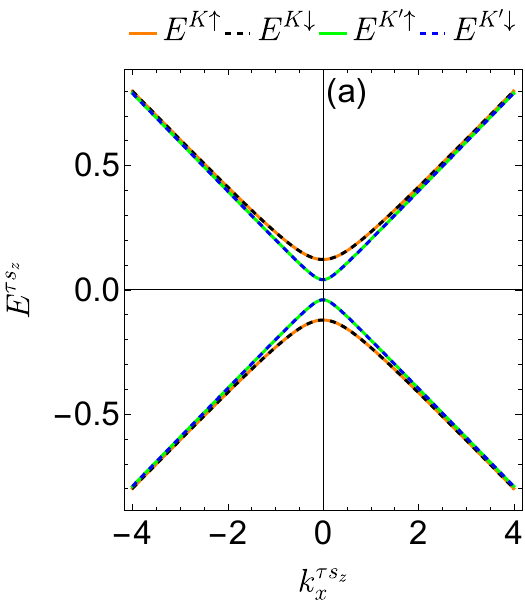}
\includegraphics[scale=0.44]{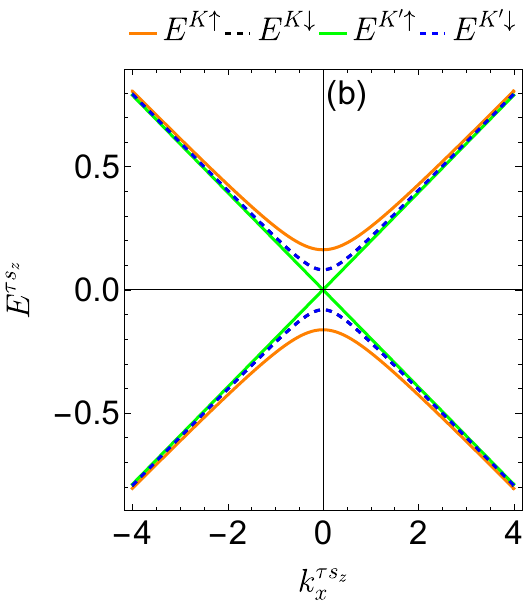}
\includegraphics[scale=0.44]{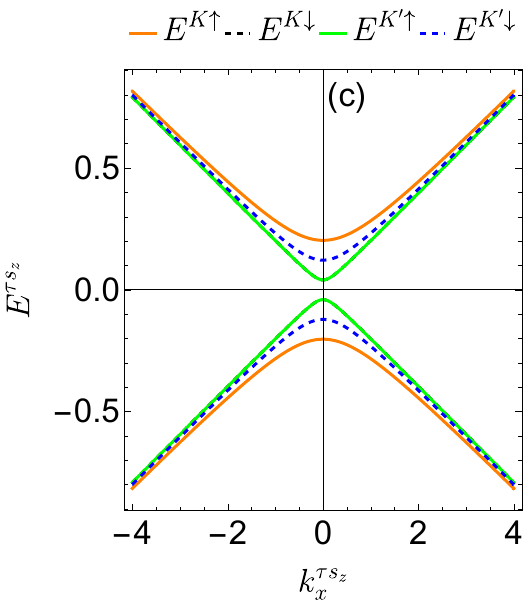}\\
\includegraphics[scale=0.44]{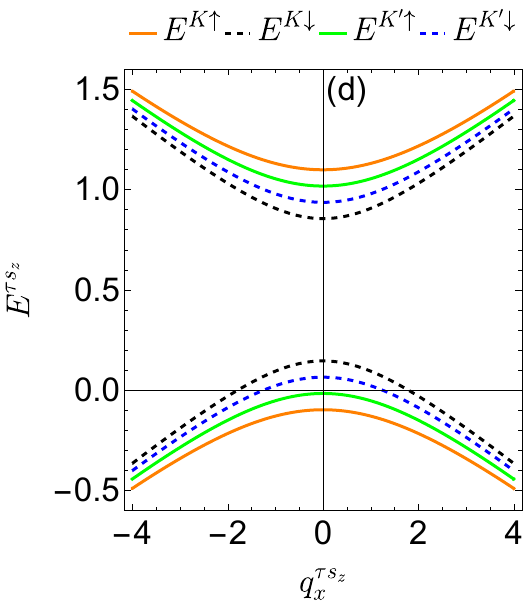}
\includegraphics[scale=0.44]{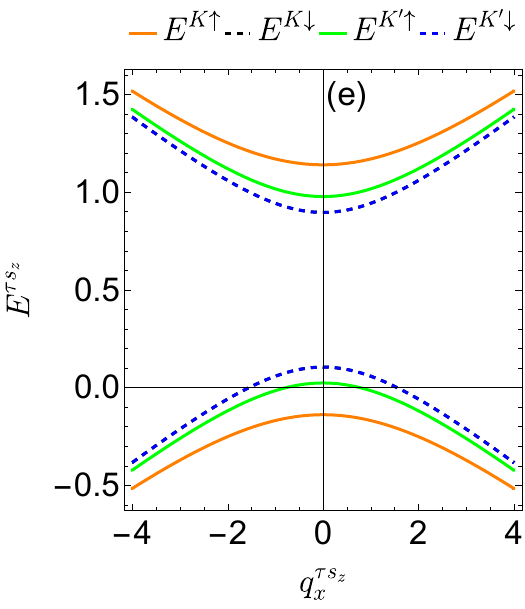}
\includegraphics[scale=0.44]{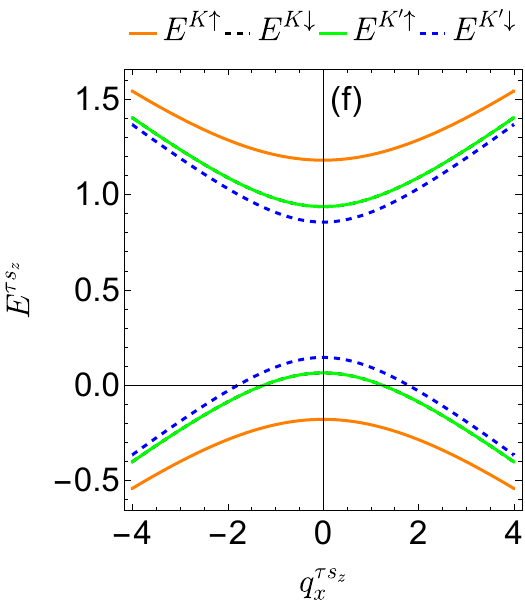}
\caption{Spin- and valley-resolved energy dispersion $E^{\eta s_z}$ of monolayer Pt$_2$HgSe$_3$ as a function of the longitudinal wave-vector component in the leads $k_x^{\eta s_z}$ (a,b,c), and in the barrier region $q_x^{\eta s_z}$ (d,e,f) for $V_0=0.5~\mathrm{eV}$, $m_s=\lambda_{\mathrm{so}}/2$, and $A_0=0.8~\mathrm{V fs/nm}$, with  (a,d): $V_z=0$, (b,e): $V_z=\lambda_{\mathrm{so}}/2$, and (c,f): $V_z=\lambda_{\mathrm{so}}$.}
\label{fig:Despersion}
\end{figure*}

To illustrate the region-dependent dispersions under the combined effects of the substrate term $V_z$ and irradiation, we plot Eq.~\eqref{eq:disper} using the lead wave vector $k_x^{\eta s_z}$ and the barrier wave vector $q_x^{\eta s_z}$ as appropriate.
Fig.~\ref{fig:Despersion} summarizes the spin- and valley-resolved dispersions that enter our tunneling problem. It compares  the undriven leads $E^{\eta s_z}(k_x^{\eta s_z})$ in panels 
(a,b,c), and the photon-dressed barrier $E^{\eta s_z}(q_x^{\eta s_z})$ in panels (d,e,f). 
We fix $V_0=0.5$~eV, $m_s=\lambda_{\mathrm{so}}/2$, and $A_0=0.8$~Vfs/nm, and tune the staggered potential from $V_z=0$ to $V_z=\lambda_{\mathrm{so}}$. The lead panels illustrate how $V_z$ reshapes the gaps at $k_x^{\eta s_z}=0$. For $V_z=0$ (Fig.~\ref{fig:Despersion}a), the spin branches are nearly degenerate within a given valley, whereas the two valleys show different gap scales. At $V_z=\lambda_{\mathrm{so}}/2$ (Fig.~\ref{fig:Despersion}b), the $\bm{K'}\uparrow$ channel closes its gap, resulting in a polarized metal state at $k_x^{\eta s_z}=0$, while the remaining valleys stay gapped. For $V_z=\lambda_{\mathrm{so}}$ (Fig.~\ref{fig:Despersion}c), the gaps reopen and the ordering of the channel gaps is rearranged. The barrier panels show what changes once the central region is driven. The spectrum is shifted by the barrier height $V_0$, and the mass terms become photon-dressed. In the effective description, $\Delta_{\mathrm{eff}}^{\eta s_z}=\Delta^{\eta s_z}+\eta\lambda_\omega$, so the light-induced contribution enters with an opposite sign in $\bm K$ and $\bm K'$. This lifts the valley degeneracy in the barrier. The exchange term further separates the  spin--resolved channels. As a result, the band edges at $q_x^{\eta s_z}=0$ split into distinct channel-dependent thresholds in Figs.~\ref{fig:Despersion}(d,e,f). The variation of $V_z$ moves these thresholds and determines which channels are closest to propagation. Overall, Fig.~\ref{fig:Despersion} sets the stage for the transport results discussed in the next section. The channel-dependent barrier gaps determine which valley has real $q_x^{\eta s_z}$ and can propagate through the barrier, and which channels are reflected. They also set the channel-dependent phase $q_x^{\eta s_z}L$ that controls Fabry--P\'erot interference. In this way, both $V_z$ and photon dressing define the key mechanism for spin-valley selectivity.

\section{Transmission and Reflection}
\label{sec:Tran_COndu}
To determine the transmission and reflection coefficients, we match the eigenspinors at the interfaces between different regions. For monolayer jacutingaite, the wave functions are constructed from spin- and valley-dependent Dirac-like Hamiltonians, leading to region-specific eigenspinors. By imposing continuity of the spinor wave functions at each interface, we obtain a system of linear equations that relates the incident, reflected, and transmitted amplitudes. Solving this system yields the transmission and reflection probabilities for each spin-valley channel. Indeed, the boundary matching is 
\begin{align}
\phi_\text{I}^{\eta s_z}(0)=\phi_\text{II}^{\eta s_z}(0),
\quad
\phi_\text{II}^{\eta s_z}(L)=\phi_\text{III}^{\eta s_z}(L).
\label{eq:BC}
\end{align}
After straightforward algebra, the spin and valley resolved transmission and reflection coefficients are obtained
\begin{widetext}
\begin{align}
 t^{\eta s_z} &=
 \frac{
 2 i \alpha^{\eta s_z}\beta^{\eta s_z}\cos\theta^{\eta s_z}\cos\phi^{\eta s_z}e^{-i k_x^{\eta s_z} L}
 }{
 2\left(
 2 i \alpha^{\eta s_z} \beta^{\eta s_z} \cos(L q_x^{\eta s_z}) \cos\theta^{\eta s_z} \cos\phi^{\eta s_z}
 +\sin(L q_x^{\eta s_z})
 \left[
 (\alpha^{\eta s_z})^2+(\beta^{\eta s_z})^2
 -2\alpha^{\eta s_z}\beta^{\eta s_z}\sin\theta^{\eta s_z}\sin\phi^{\eta s_z}
 \right]
 \right)
 },
 \label{eq:t}\\
 r^{\eta s_z} &=
 \frac{
 i e^{-i(L q_x^{\eta s_z}+\theta^{\eta s_z}+2\phi^{\eta s_z})}
 \left(-1+e^{2 i L q_x^{\eta s_z}}\right)
 \left(-e^{i\theta^{\eta s_z}} \alpha^{\eta s_z}+e^{i \phi^{\eta s_z}} \beta^{\eta s_z}\right)
 \left(\alpha^{\eta s_z}+e^{i(\theta^{\eta s_z}+\phi^{\eta s_z})}\beta^{\eta s_z}\right)
 }{
 2\left(
 2 i \alpha^{\eta s_z} \beta^{\eta s_z} \cos(L q_x^{\eta s_z}) \cos\theta^{\eta s_z} \cos\phi^{\eta s_z}
 +\sin(L q_x^{\eta s_z})
 \left[
 (\alpha^{\eta s_z})^2+(\beta^{\eta s_z})^2
 -2\alpha^{\eta s_z}\beta^{\eta s_z}\sin\theta^{\eta s_z}\sin\phi^{\eta s_z}
 \right]
 \right)
 }
 \label{eq:r}.
\end{align}
\end{widetext}

To obtain the transmission and reflection probabilities, we use the longitudinal current density. They are defined as the ratios of the transmitted and reflected currents to the incident current, ensuring proper normalization of the scattering states. The current density is evaluated from the expectation value of the velocity operator with respect to the eigenspinors in each region. By applying this procedure to all spin-valley channels, we extract the channel-resolved transmission and reflection probabilities that govern the transport properties of the system. Consequently, for our model, we have
\begin{align}
J^{\eta s_z}_x=-e v_F(\phi^{\eta s_z})^\dagger \sigma_x\phi^{\eta s_z}
\label{eq:current}
\end{align}
The relations $T^{\eta s_z}
=\frac{|J_x^{\mathrm{tr}}|}{|J_x^{\mathrm{in}}|}$, and $R^{\eta s_z}
=\frac{|J_x^{\mathrm{re}}|}{|J_x^{\mathrm{in}}|}$, 
allow us to obtain $T^{\eta s_z} =|t^{\eta s_z}|^2$, and $ R^{\eta s_z}
=|r^{\eta s_z}|^2$, respectively. Explicitly, we have
\begin{widetext}
\begin{align}
	&		R^{\eta s_z}=
			\frac{1}{2} \frac{\sin^2(q_x^{\eta s_z} L ) ((\frac{(\alpha^{\eta s_z})^2 + (\beta^{\eta s_z})^2}{\alpha^{\eta s_z} \beta^{\eta s_z}})^2 - \cos2\phi^{\eta s_z}-\cos2\theta^{\eta s_z}- \frac{(\alpha^{\eta s_z})^2 + (\beta^{\eta s_z})^2}{\alpha^{\eta s_z} \beta^{\eta s_z}}\sin\phi^{\eta s_z}\sin\theta^{\eta s_z})}
			{ \cos^2\phi^{\eta s_z}\cos^2\theta^{\eta s_z}\cos^2(q_x^{\eta s_z} L)	+ \sin^2(q_x^{\eta s_z} L )(\frac{(\alpha^{\eta s_z})^2 + (\beta^{\eta s_z})^2}{2\alpha^{\eta s_z} \beta^{\eta s_z}} -   \sin\phi^{\eta s_z}\sin\theta^{\eta s_z})^2}.
		\\
        &
			T^{\eta s_z}=
			  \frac{\cos^2\phi^{\eta s_z}\cos^2\theta^{\eta s_z}}
			{ \cos^2\phi^{\eta s_z}\cos^2\theta^{\eta s_z}\cos^2(q_x^{\eta s_z} L)	+ \sin^2(q_x^{\eta s_z} L )(\frac{(\alpha^{\eta s_z})^2 + (\beta^{\eta s_z})^2}{2\alpha^{\eta s_z} \beta^{\eta s_z}} -  \sin\phi^{\eta s_z}\sin\theta^{\eta s_z})^2}.
		\end{align}
        \end{widetext}
We verify the conservation for propagating modes through $T^{\eta s_z}+R^{\eta s_z}=1$.
Due to the mirror symmetry of the junction, the transmission satisfies $T^{\eta s_z}(k_y)=T^{\eta s_z}(-k_y)$.
At normal incidence ($k_y=0$) and without irradiation ($\lambda_\omega=0$), perfect transmission occurs when the Dirac mass vanishes in the incident channel. In this case the spinor mismatch disappears and one finds $\alpha^{\eta s_z}=\beta^{\eta s_z}=1$. As a result,  the barrier becomes transparent, and $T^{\eta s_z}=1$, reflecting  Klein tunneling~\cite{CastroNeto2009,Hassane2022}. In our model, the mass-cancellation condition reads $\Delta^{\eta s_z}=0$, with $\Delta^{\eta s_z}$ given by Eq.~\eqref{eq:gap}. Therefore, Klein tunneling is channel-selective and can be tuned by $m_s$ and $V_z$. For example, choosing $m_s=V_z=\lambda_{\mathrm{so}}/2$ yields $\Delta^{K'\uparrow}=0$ and therefore perfect transmission in the $K'\uparrow$ channel. Other parameter choices can single out other channels through the same condition.

\begin{figure}[ht!]
\centering 
\includegraphics[scale=0.44]{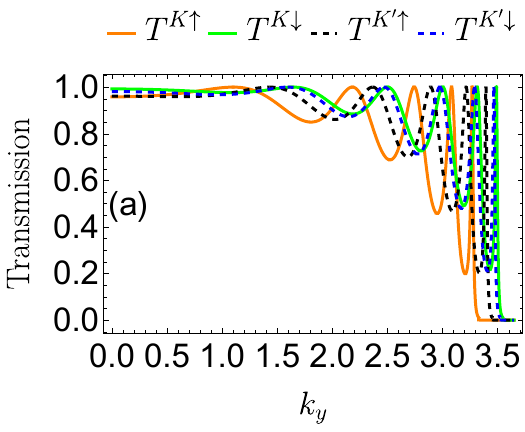}
\includegraphics[scale=0.44]{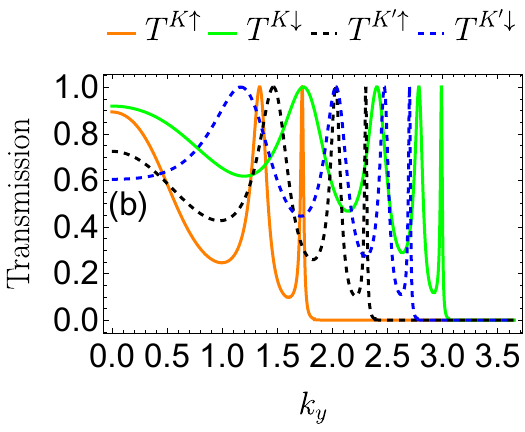}\\
\includegraphics[scale=0.44]{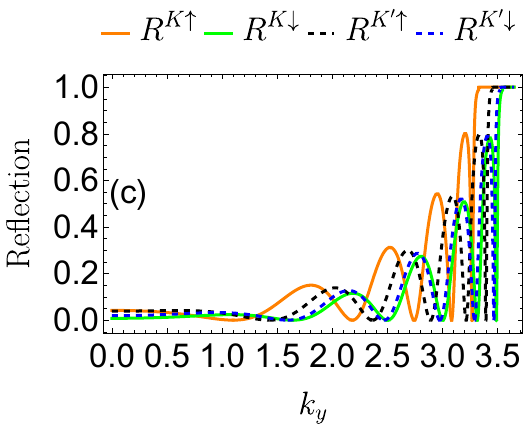}
\includegraphics[scale=0.44]{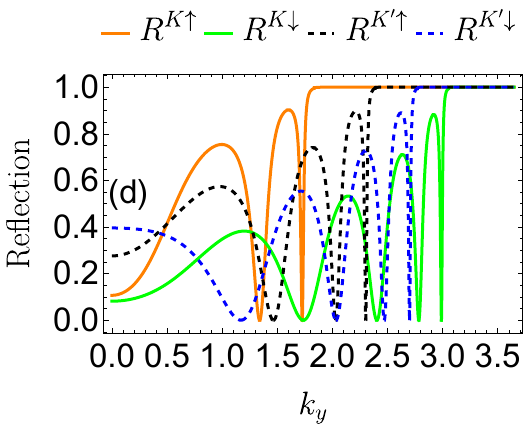}
\caption{Spin- and valley-resolved transmission $T^{\eta s_z}$ (a,b) and reflection $R^{\eta s_z}$ (c,d) probabilities as functions of the transverse wave vector $k_y$ for $V_0 = 0.5$~eV, $L = 5$~nm, $E^{\eta s_z} = 1.2$~eV, $m_s = \lambda_{\mathrm{so}}/2$ and $V_z = 0$. (a,c): Moderate driving amplitude $A_0 = 0.4$~Vfs/nm, and (b,d): strong driving amplitude $A_0 = 0.8$~V fs/nm.}
\label{fig:TRky}
\end{figure}

Figure~\ref{fig:TRky} shows the spin- and valley-resolved tunneling probabilities, $T^{\eta s_z}(k_y)$ and $R^{\eta s_z}(k_y)$, as functions of the conserved transverse momentum $k_y$. The value of $k_y$ fixes the incident angle, and also enters the boundary matching. Thus, it controls the interface mismatch and the interference condition set by the barrier phase. We first consider moderate driving in Figs.~\ref{fig:TRky}(a,c), with $A_0=0.4~\mathrm{Vfs/nm}$. Near normal incidence, all channels transmit strongly, and as $k_y$ is increased
two regimes appear. For small and intermediate $k_y$, the barrier mode is propagating. In this case, $q_x^{\eta s_z}$ is real, and the transmission develops Fabry--P\'erot oscillations 
{governed by the phase} $q_x^{\eta s_z}L$, with maxima near $q_x^{\eta s_z}L \simeq n\pi$. For larger $k_y$, the barrier becomes evanescent. The crossover is set by $(E^{\eta s_z}-V_0)^2 < (\Delta_{\mathrm{eff}}^{\eta s_z})^2 + (\hbar v_F k_y)^2$, for which $q_x^{\eta s_z}$ in Eq.~\eqref{eq:qx} becomes imaginary. In this regime, $T^{\eta s_z}$ collapses, and $R^{\eta s_z}$ approaches unity. We now increase the driving strength in Figs.~\ref{fig:TRky}(b,d), with $A_0=0.8~\mathrm{Vfs/nm}$. The correction to the effective mass term $\Delta_{\mathrm{eff}}^{\eta s_z}=\Delta^{\eta s_z}+\eta\lambda_\omega$ in Eq.~\eqref{eq:effgap}, caused by the Floquet dressing, has a stronger effect since $\lambda_\omega\propto A_0^2$, and has the opposite sign in $\bm K$ and $\bm K'$. As a result, both the propagating $k_y$ window and the phase $q_x^{\eta s_z}L$ become strongly channel dependent. One then finds finite $k_y$ intervals where a selected spin-valley channel remains transmitting while competing channels are suppressed, thus, the contrast is enhanced. In this sense, Fig.~\ref{fig:TRky} demonstrates that the irradiated barrier acts as a tunable spin-valley filter that is controlled by the photon-dressed mass $\Delta_{\mathrm{eff}}^{\eta s_z}$.

\begin{figure}[ht!]
\centering 
\includegraphics[scale=0.44]{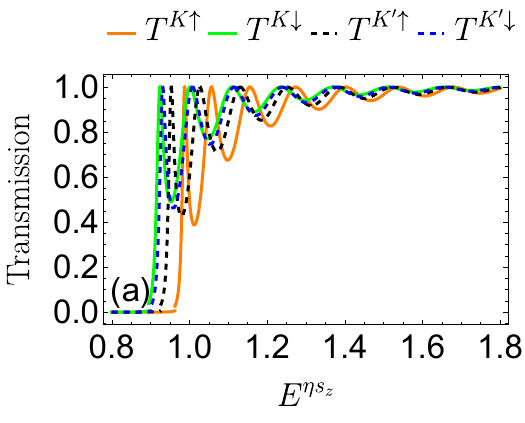}
\includegraphics[scale=0.44]{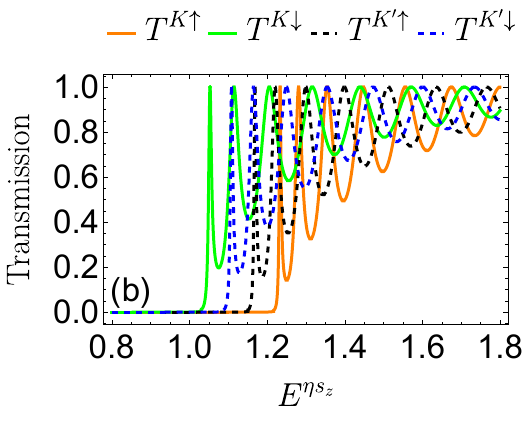}\\
\includegraphics[scale=0.44]{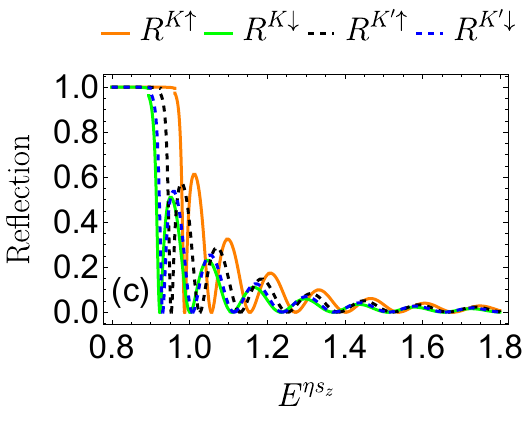}
\includegraphics[scale=0.44]{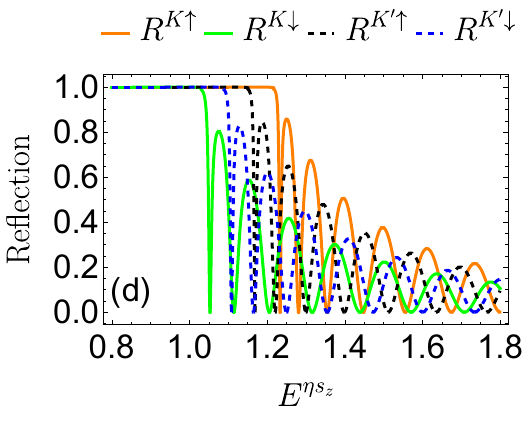}
\caption{Spin- and valley-resolved  transmission $T^{\eta s_z}$ (a,b) and  reflection $R^{\eta s_z}$  probabilities as functions of the incident energy $E^{\eta s_z}$ for $V_0 = 0.5$~eV, $L = 4$~nm, $k_y = 2$~nm$^{-1}$, $m_s = \lambda_{\mathrm{so}}/2$, and $V_z = 0$. (a,c): Moderate driving amplitude $A_0 = 0.4$~Vfs/nm, and (b,d): strong driving amplitude $A_0 = 0.8$~V fs/nm.}
\label{fig:TREne}
\end{figure}

In Fig.~\ref{fig:TREne}, we fix an oblique incidence $k_y=2~\mathrm{nm^{-1}}$, and sweep the incident energy to show the  behavior of $T^{\eta s_z}(E^{\eta s_z})$ and $R^{\eta s_z}(E^{\eta s_z})$. This scan is useful because it isolates the propagation thresholds. Sweeping $E^{\eta s_z}$ directly tunes the longitudinal barrier wave vector $q_x^{\eta s_z}(E^{\eta s_z})$ in Eq.~\eqref{eq:qx}. For moderate driving in Figs.~\ref{fig:TREne}(a,c), with $A_0=0.4~\mathrm{Vfs/nm}$, at low energies, the resulting mode is evanescent, and almost all channels are reflected, resulting in an opaque barrier. As $E^{\eta s_z}$ is increased, $q_x^{\eta s_z}$ becomes real, and the channels turn on sequentially. The threshold energies are close, but they are not identical since the Floquet correction is weak, which results in a small renormalization of the effective band gap $\Delta_{\mathrm{eff}}^{\eta s_z}=\Delta^{\eta s_z}+\eta\lambda_\omega$ in Eq.~\eqref{eq:effgap}. For the parameters shown, the $K\uparrow$ channel turns on last. Once a channel is propagating, the finite barrier produces Fabry--P\'erot oscillations, with the relevant phase  $q_x^{\eta s_z}L$. For stronger driving in Figs.~\ref{fig:TREne}(b,d), with $A_0=0.8~\mathrm{Vfs/nm}$, the same picture emerges, but more sharply as the dressing is stronger, resulting in a larger effective gap, thus more incident energy is required to activate the transmission. With exchange, the spin channels are also separated, and therefore, the thresholds are no longer clustered; they split into clearly distinct onsets. As a result, one obtains an extended energy window where a single channel dominates the transmitted current while the others remain essentially reflected. This appears as long plateaus near unity, with sharp steps at the corresponding onsets. As a result, Fig.~\ref{fig:TREne}, therefore, demonstrates threshold filtering in energy.

\begin{figure}[ht!]
\centering 
\includegraphics[scale=0.44]{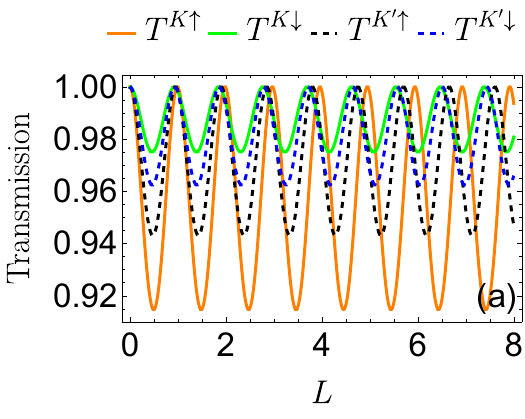}
\includegraphics[scale=0.44]{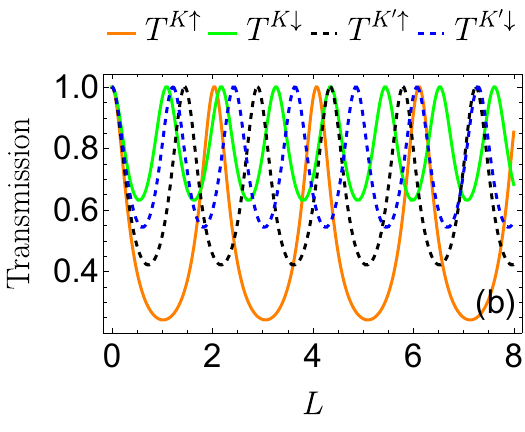}\\
\includegraphics[scale=0.44]{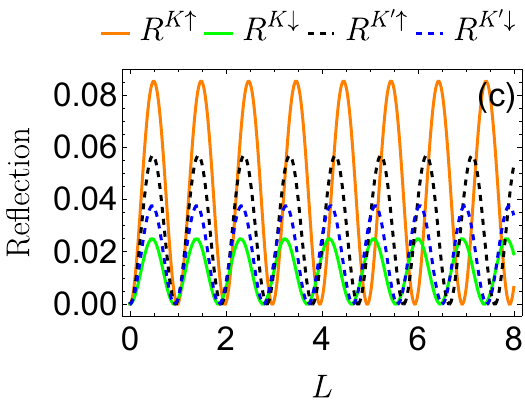}
\includegraphics[scale=0.44]{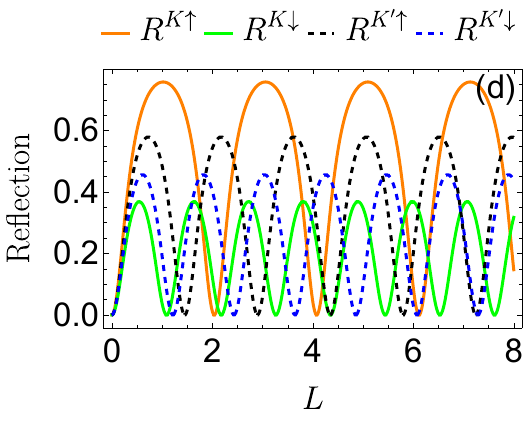}
\caption{Spin- and valley-resolved transmission $T^{\eta s_z}$ (a,b) and reflection $R^{\eta s_z}$ (c,d) probabilities as functions of the barrier width $L$ for $V_0 = 0.5$~eV, $E^{\eta s_z} = 1.2$~eV, $k_y = 1$~nm$^{-1}$, $m_s = \lambda_{\mathrm{so}}/2$ and $V_z = 0$. (a,c): Moderate driving amplitude $A_0 = 0.4$~Vfs/nm, and (b,d): strong driving amplitude $A_0 = 0.8$~V fs/nm.}
    \label{fig:TRL}
\end{figure}


 Figure~\ref{fig:TRL} shows the spin- and valley-resolved tunneling probabilities $T^{\eta s_z}(L)$ and $R^{\eta s_z}(L)$ as functions of the barrier width $L$, isolating the interference effects.
However, this is not easily seen from the $k_y$ and $E^{\eta s_z}$ sweeps in Figs.~\ref{fig:TRky} and \ref{fig:TREne}, since varying $L$ does not change whether a channel is propagating or evanescent. It mainly changes the phase accumulated in the barrier. When $q_x^{\eta s_z}$ is real, the relevant phase is $q_x^{\eta s_z}L$, and the junction behaves as a Fabry--P\'erot cavity. Transmission oscillates with a period set by the resonance condition $q_x^{\eta s_z}L\simeq n\pi$. The reflection follows in a complementary way: every transmission peak corresponds to a reflection minimum, and vice versa, and $T^{\eta s_z}+R^{\eta s_z}=1$ is valid for any given $L$. For a weaker drive in Figs.~\ref{fig:TRL}(a,c), with $A_0=0.4~\mathrm{Vfs/nm}$, all channels remain highly transmitting over the plotted range of $L$. The oscillation period is small and is channel-dependent. The clearest modulation appears in the $K\uparrow$ channel, consistent with the largest mismatch for this set of parameters. For a stronger drive in Figs.~\ref{fig:TRL}(b,d), with $A_0=0.8~\mathrm{Vfs/nm}$, the same picture remains, but the oscillation period increases. As discussed earlier, the driving is stronger, resulting in a larger $\Delta_{\mathrm{eff}}$ since $\lambda_\omega\propto A_0^2$. As a result, the resonance condition $q_x^{\eta s_z}L\simeq n\pi$ is met at just a few width values for different channels. One then finds $L$ intervals where a selected channel sits close to resonance, while others are off resonance. The same effect appears in reflection, which peaks at transmission minima. Fig.~\ref{fig:TRL}, therefore, demonstrates an explicit geometric control of spin-valley selectivity in the propagating regime, and shows how the drive enhances this effect by increasing the channel dependence of $q_x^{\eta s_z}$.

\begin{figure}[ht!]
\centering 
\includegraphics[scale=0.34]{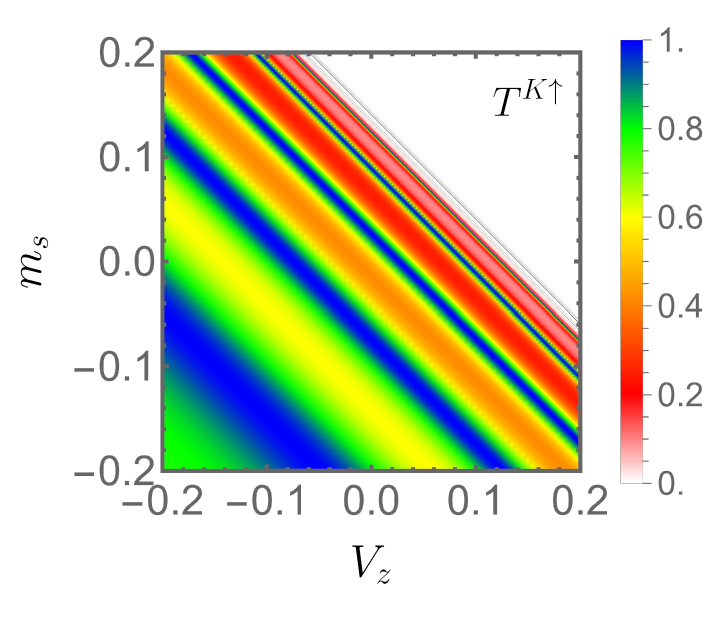}
\includegraphics[scale=0.34]{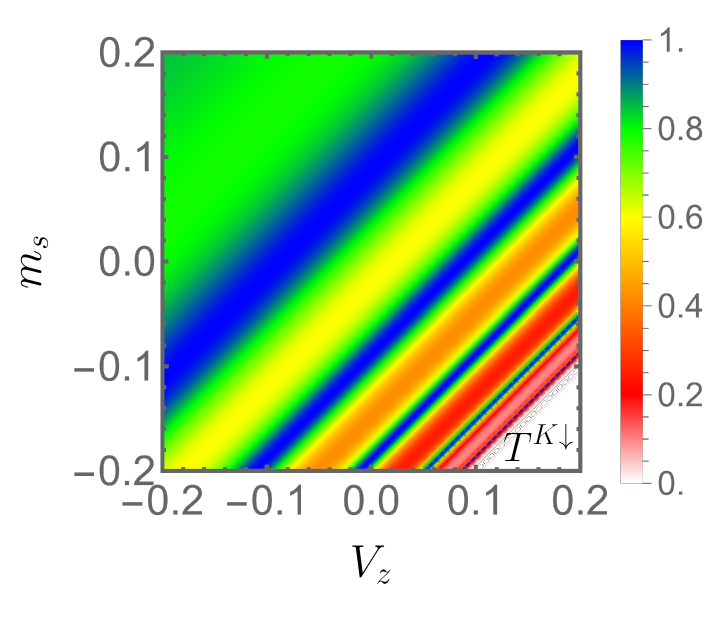}\\
\includegraphics[scale=0.34]{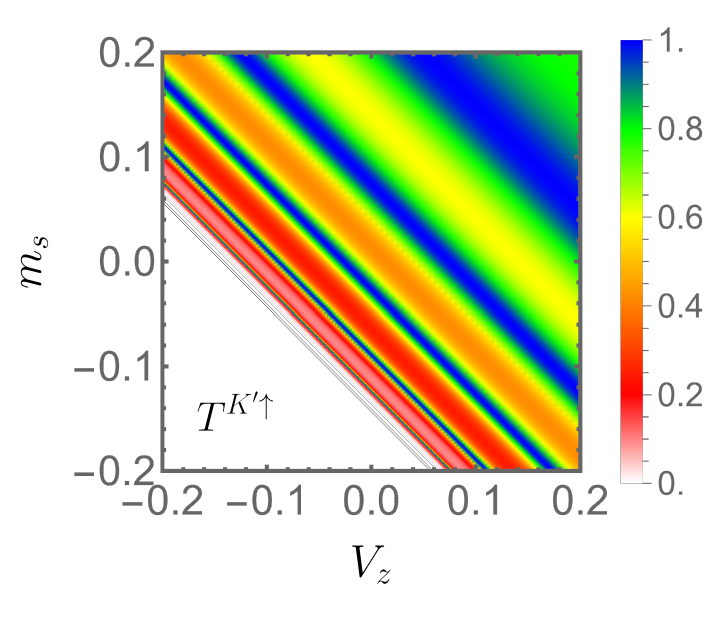}
\includegraphics[scale=0.34]{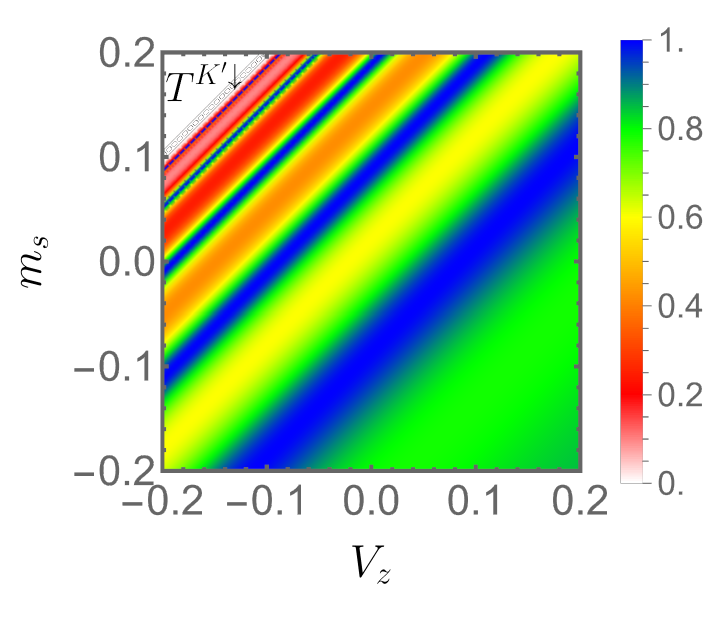}
\caption{Density plot of spin- and valley-resolved transmission  probability $T^{\eta s_z}$ as functions of the substrate mass $m_s$ and the sublattice staggered potential $V_z$ for $V_0 = 0.5$~eV, $E^{\eta s_z} = 1.2$~eV, $k_y = 0.5$~nm$^{-1}$, $L = 5$ nm, and $A_0 = 0.8$~Vfs/nm.}
    \label{fig:TVzms}
\end{figure}

Figure~\ref{fig:TVzms} shows the spin- and valley-resolved transmission $T^{\eta s_z}$ as a function of the barrier knobs $m_s$ and $V_z$ for strong driving $A_0=0.8$~V$\cdot$fs/nm. In this scan, we analyze how $m_s$ and $V_z$ alter the transport because they enter the barrier mass term through Eq.~\eqref{eq:gap}. Therefore, they tune the longitudinal barrier wave vector $q_x^{\eta s_z}$ in Eq.~\eqref{eq:qx}. In this way, sweeping $(m_s, V_z)$ directly reveals where a given channel is propagating in the barrier and where it becomes evanescent. We begin with the regions of finite transmission. There, the barrier mode is propagating, and $q_x^{\eta s_z}$ is real. The barrier then produces Fabry--P\'erot oscillations, appearing as diagonal fringes of alternating high and low transmission. The relevant phase is $q_x^{\eta s_z}L$, and then  the fringes track the resonance condition $q_x^{\eta s_z}L\simeq n\pi$. Varying $(m_s,V_z)$ shifts $\Delta_{\mathrm{eff}}^{\eta s_z}$ in Eq.~\eqref{eq:effgap}, which shifts $q_x^{\eta s_z}$ and moves these fringes. The orientation of the fringes flips between spin-up and spin-down channels. This is expected, since $m_s$ enters the mass term with the prefactor $s_z$. A second regime appears as extended low-transmission regions. In these regions, the barrier mode is non-propagating, $q_x^{\eta s_z}$ becomes imaginary, and $T^{\eta s_z}$ collapses. The location of the suppressed channels is strongly valley-dependent. For spin-up, $K\uparrow$ is suppressed in the $(m_s>0,V_z>0)$ quadrant, while $K'\uparrow$ is suppressed in the opposite $(m_s<0,V_z<0)$ quadrant. For spin down, the suppression shifts to the opposite sign quadrants: $K\downarrow$ is suppressed for $(m_s<0, V_z>0)$, while $K'\downarrow$ is suppressed for $(m_s>0, V_z<0)$. This systematic shift is the transport signature of the drive-induced valley term, which enters with an opposite sign in $\bm K$ and $\bm K'$. Taken together, Fig.~\ref{fig:TVzms} provides an operating map. By tuning $(m_s,V_z)$, one can place a target spin-valley channel in a propagating, high-transmission fringe while pushing competing channels into suppressed regimes, offering an efficient knob for spin-valley filtering.

\section{Conductance}\label{CCC}

We compute the zero-temperature conductance per unit width of the monolayer jacutingaite junction within the Landauer–Büttiker framework \cite{PhysRevB.103.245432,Qiu2020,Li2025SciRep,Nguyen2011JAP}. In this approach, the conductance is obtained by summing over all transverse modes that contribute to transport at a given Fermi energy. Specifically, the spin- and valley-resolved conductance is expressed as
 \begin{align}
G_{\eta s_z}(E_F)=\frac{e^2}{h}\int\frac{dk_y}{2\pi}T^{\eta s_z}(E_F,k_y),
 \end{align}
where the integration over the conserved transverse momentum $k_y$ is restricted to propagating incident states in the leads. This restriction ensures that only modes with real longitudinal wave vector $k_x^{\eta s_z}$ contribute to the current. The transmission probability $T^{\eta s_z}(E_F,k_y)$ encodes the full information about the scattering processes across the irradiated barrier, including the effects of the photon-dressed mass, substrate-induced potential $V_z$, and exchange field $m_s$. 
The conductance demonstrates the interaction between band structure engineering and quantum interference. 
{The Fermi energy fixes how many transverse modes are available, while $k_y$ fixes how strongly each is scattered, and the integration averages the resulting angular interference pattern.} 
What survives this average is the channel dependence of $\Delta_{\mathrm{eff}}^{\eta s_z}$. By Eq.~\eqref{eq:qx} a channel propagates only for $|k_y|<\sqrt{(E^{\eta s_z}-V_0)^2-(\Delta_{\mathrm{eff}}^{\eta s_z})^2}/\hbar v_F$, so its window narrows as its barrier mass grows. Far above threshold, the four windows are wide and nearly equal, and the sector conductances are therefore similar. Near threshold, however, they differ, and the conductance becomes spin- and valley-selective. The system reaches states in which one spin-valley channel controls all transport, leading to a highly polarized current. 

\begin{figure}[ht!]
\centering 
\includegraphics[scale=0.46]{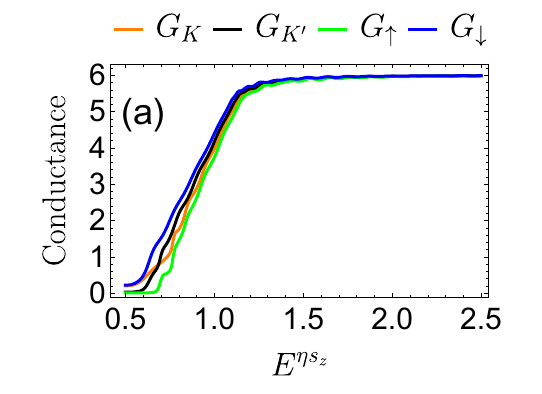}
\includegraphics[scale=0.46]{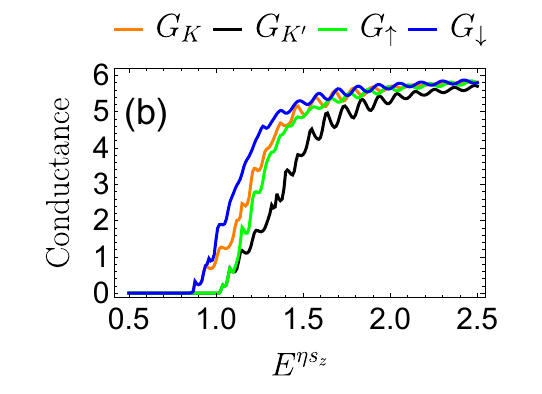}\\
\includegraphics[scale=0.46]{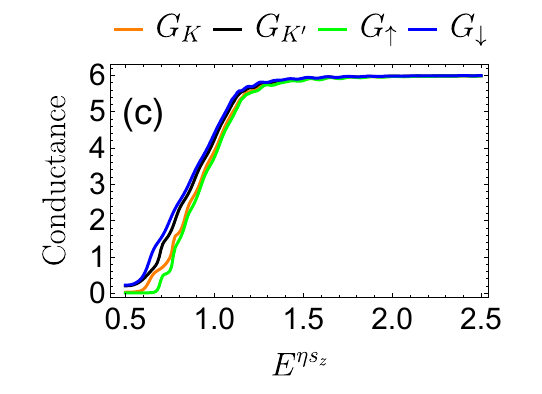}
\includegraphics[scale=0.46]{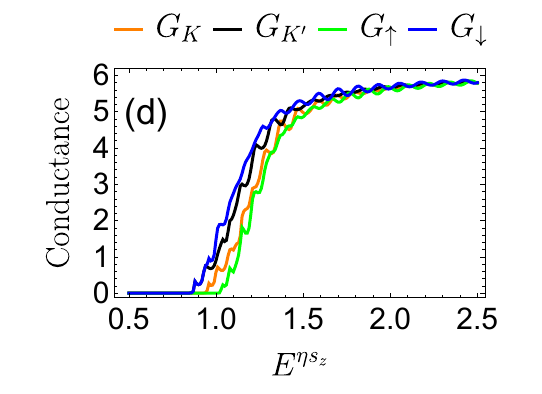}
\caption{Valley- and spin-resolved conductance $G_{\eta s_z}$ as a function of incident energy $E^{\eta s_z}$ for $V_0 = 0.5$~eV, $L = 5$~nm. (a,b):~$m_s = \lambda_{\mathrm{so}}/2$,  $V_z = 0$, and (c,d): $m_s = 0$, $V_z = \lambda_{\mathrm{so}}/2$. (a,c): Moderate driving amplitude $A_0 = 0.4$~Vfs/nm, and  (b,d): strong driving amplitude $A_0 = 0.8$~Vfs/nm.}
    \label{fig:CondEner}
\end{figure}

Figure~\ref{fig:CondEner} converts the channel-resolved transmission in Figs.~\ref{fig:TRky} and \ref{fig:TREne} into a directly measurable quantity. We illustrate the spin- and valley-resolved conductance $G_{\eta s_z}$ as a function of the incident energy $E^{\eta s_z}$. Increasing $E^{\eta s_z}$ increases the number of propagating transverse modes. As a result, $G(E^{\eta s_z})$ increases once propagating channels open by satisfying the inequality $(E^{\eta s_z}-V_0)^2 >(\Delta_{\mathrm{eff}}^{\eta s_z})^2 + (\hbar v_F k_y)^2$, resulting in a real barrier wave vector $q_x^{\eta s_z}(E^{\eta s_z})$ in Eq.~\eqref{eq:qx}. This extends the transmission threshold behavior to conductance, seen in Fig.~\ref{fig:TREne}. Weak ripples are the traces of Fabry-P\'erot interference resulting from the $k_y$ integration. We begin with the exchange-driven configuration, $m_s=\lambda_{\mathrm{so}}/2$ and $V_z=0$ in Figs.~\ref{fig:CondEner}(a,b). For moderate driving in Fig.~\ref{fig:CondEner}a, the conductance traces nearly overlap, and both spin and valley selectivity are weak. For strong driving in Fig.~\ref{fig:CondEner}b, the same picture becomes sharper. The traces separate, and a clear splitting develops between $\bm K$ and $\bm K'$ valleys, while the spin splitting becomes visible as well. The origin is the stronger photon-dressed mass shift in Eq.~\eqref{eq:effgap}, which shifts the effective thresholds with opposite sign in the two valleys, leading to a larger effective gap and requiring higher energy to reach the propagating regime, seen in Fig.~\ref{fig:TREne}. Since $G_{\bm{K}} > G_{\bm{K'}}$, a positive valley polarization is expected in this case. In the absence of exchange $m_s = 0$ and $V_z = \lambda_{\mathrm{so}}/2$ in Fig.~\ref{fig:CondEner}(c,d), the spin contrast is more pronounced since $G_{\bm{K'}} > G_{\bm{K}}$, resulting in negative valley polarization $P_v$. Overall, Fig.~\ref{fig:CondEner} shows that the threshold filtering of Fig.~\ref{fig:TREne} persists at the conductance level. It also shows that strong driving is the key ingredient for robust spin and valley contrast.

\begin{figure}[ht!]
\centering 
\includegraphics[scale=0.46]{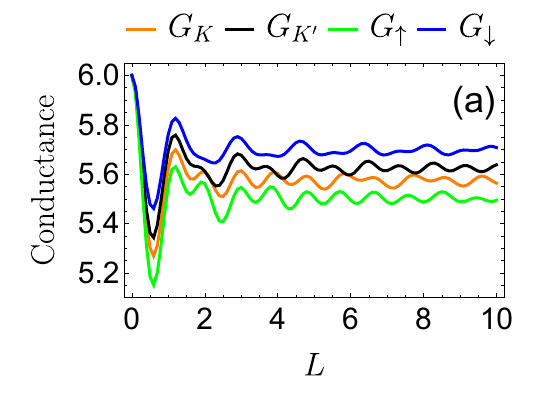}
\includegraphics[scale=0.46]{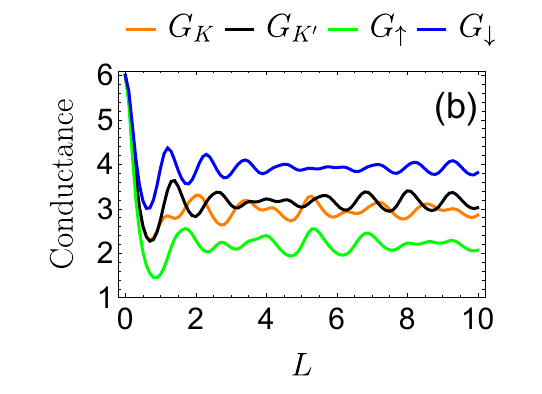}\\
\includegraphics[scale=0.46]{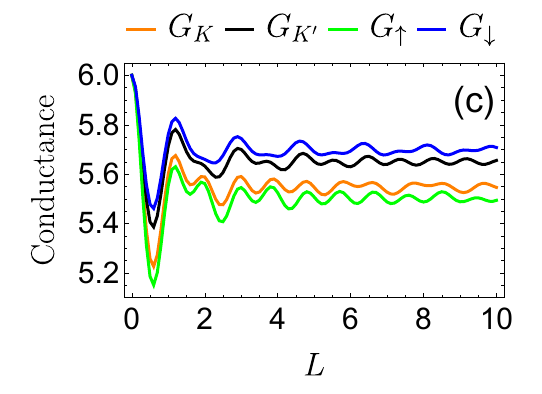}
\includegraphics[scale=0.46]{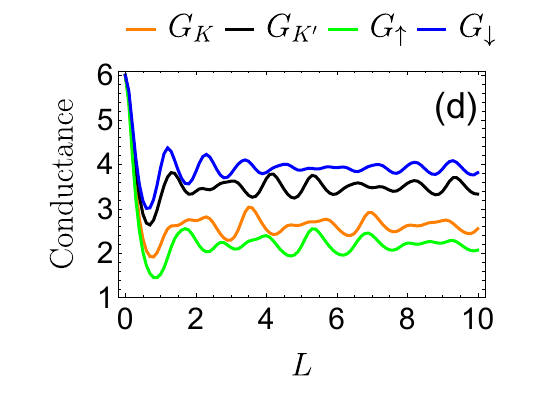}
\caption{Valley- and spin-resolved conductance $G_{\eta s_z}$ as a function of the barrier width $L$ for $V_0 = 0.5$~eV, $E^{\eta s_z} = 1.2$ eV, (a,b): $m_s = \lambda_{\mathrm{so}}/2$,  $V_z = 0$, and  (c,d): $V_z = \lambda_{\mathrm{so}}/2$, $m_s = 0$. (a,c): Moderate driving amplitude $A_0 = 0.4$~Vfs/nm, and (b,d): strong driving amplitude $A_0 = 0.8$~Vfs/nm.}
   \label{fig:CondL}
\end{figure}

Figure~\ref{fig:CondL} shows the conductance $G_{\eta s_z}$ as a function of the barrier width $L$. Thus far, we have focused on tuning with mass knobs, but here we examine the junction geometrically. Only region II is photon dressed, so varying $L$ changes how long carriers propagate under the dressed Dirac mass. In the effective description, $\Delta_{\mathrm{eff}}^{\eta s_z}=\Delta^{\eta s_z}+\eta\lambda_\omega$ in Eq.~\eqref{eq:effgap}, and the longitudinal wave vector $q_x^{\eta s_z}$ is channel dependent. Each channel therefore accumulates a different phase $q_x^{\eta s_z}L$. This produces oscillations in $G_{\eta s_z}(L)$ and provides a geometric optimization knob. By choosing $L$, one can place a target channel near resonance while pushing the competing channels toward off-resonance. At moderate driving in Figs.~\ref{fig:CondL}(a,c), with $A_0=0.4~\mathrm{Vfs/nm}$, the photon-dressed term is weak, so the resulting channel dependence of $q_x^{\eta s_z}$ is weak. After the $k_y$ integration, the conductance traces remain close, and the selectivity is limited. At stronger driving in Figs.~\ref{fig:CondL}(b,d), with $A_0=0.8~\mathrm{V\cdot fs/nm}$, $\lambda_\omega\propto A_0^2$ becomes large and carries the valley index $\eta$. The valleys, therefore, separate, and $G_{K}$ and $G_{K'}$ split over essentially the full $L$ range. With exchange present, the spin channels separate as well, and $G_{\uparrow}$ and $G_{\downarrow}$ split. In this regime, the variation of $L$ becomes an efficient way to optimize the conductance of a chosen channel. By comparing panels (a,b) to (c,d), we isolate the role of the exchange $m_s$. For $m_s\neq 0$ in panels (a,b), strong driving yields valley and spin discrimination. For $m_s=0$ but $V_z\neq 0$ in panels (c,d), valley selectivity remains, but the spin contrast is reduced. In this sense, the photon dressing sets the valley contrast. Therefore, Fig.~\ref{fig:CondL} shows that the exchange $m_s$ sets the spin contrast, and $L$ provides the geometric knob that optimizes the conductance of the desired channel. In addition, we observe that $G_{\uparrow}$ is consistently smaller than $G_{\downarrow}$ in all panels, implying a negative spin polarization $P_s$. By contrast, the relative magnitude of $G_{K}$ and $G_{K'}$ depends on the parameters and can interchange, so that the valley polarization $P_v$ changes sign, as discussed in Sec.~\ref{sec:Polarization}.

\begin{figure}[ht!]
\centering 
\includegraphics[scale=0.34]{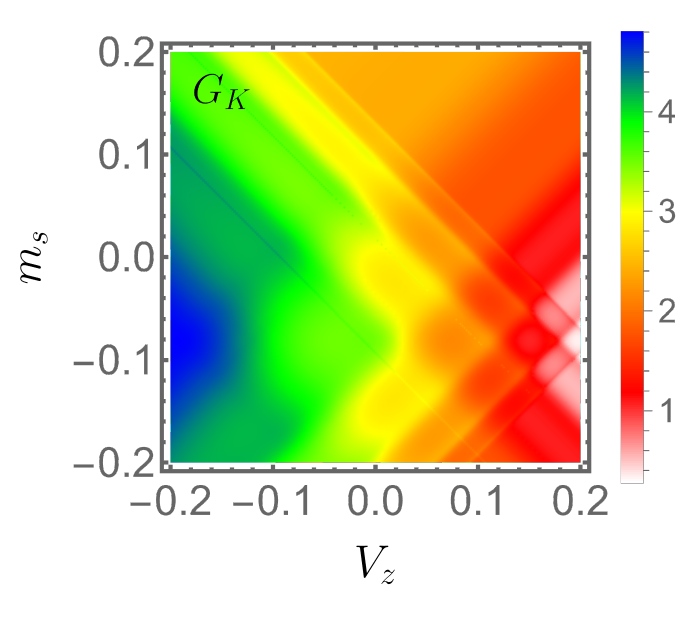}
\includegraphics[scale=0.34]{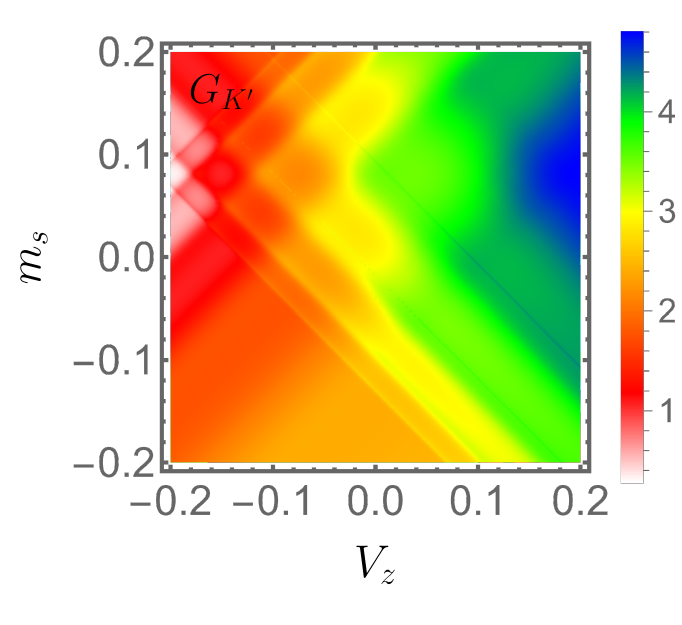}\\
\includegraphics[scale=0.34]{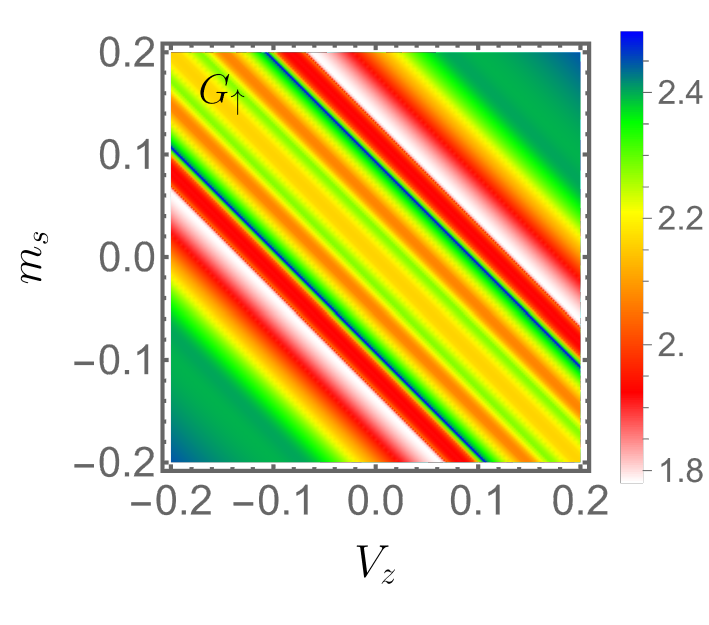}
\includegraphics[scale=0.34]{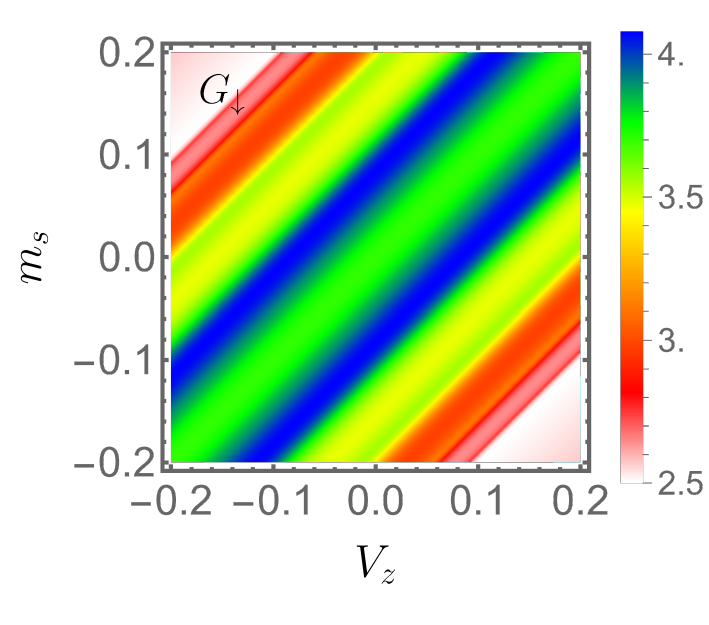}
\caption{Density plot of spin- and valley-resolved conductance $G_{\eta s_z}$ as functions of the substrate mass $m_s$ and the sublattice staggered potential $V_z$ for $V_0 = 0.5$~eV, $E^{\eta s_z} = 1.2$~eV, $L = 5$ nm, and $A_0 = 0.8$~V$\cdot$fs/nm.}
    \label{fig:GVzms}
\end{figure}

Figure~\ref{fig:GVzms} presents the conductance-level counterpart to Fig.~\ref{fig:TVzms}. It shows valley- and spin-resolved conductances as functions of the two static barrier knobs, $m_s$ and $V_z$. The fine transmission fringes of $T^{\eta s_z}$ are averaged out, and the remaining structure is therefore the robust one. We first focus on the valley panels, $G_{K}$ and $G_{K'}$, which are strongly complementary. Regions of large $G_{K}$ coincide with suppressed $G_{K'}$, and vice versa. This behavior follows from photon dressing in the barrier. The drive-induced valley term enters with an opposite sign in $\bm K$ and $\bm K'$. Therefore, tuning $(m_s, V_z)$ to place one valley in a conducting regime tends to push the other valley into a suppressed one. In this sense, the map provides an operating chart. Flipping the sign combination of $(m_s, V_z)$ switches the transmitted valley. A similar level of structure appears in the spin panels, $G_{\uparrow}$ and $G_{\downarrow}$. We find broad diagonal regions of enhanced and suppressed conductance, as expected. The exchange mass enters the barrier mass term with opposite sign for opposite spins. Importantly, the diagonal structure remains after the $k_y$ integration. This indicates that the spin selectivity is not a fine-interference effect, but rather survives at the conductance level. Taken together, Fig.~\ref{fig:GVzms} identifies operating regimes with strong valley contrast driven by photon dressing and strong spin contrast driven by exchange. It also motivates the polarization analysis in Sec.~\ref{sec:Polarization}.
 
\section{Polarization measures}
\label{sec:Polarization}

To better understand the transport properties in monolayer jacutingaite (Pt$_2$HgSe$_3$), we focus on the study of polarization under the influence of the substrate term $s_z m_{s} \sigma_z$, the sublattice staggered potential $V_z$, and the barrier width $L$. From the spin- and valley-resolved channel conductances, we define the spin and valley polarizations as \cite{PhysRevB.103.245432,Qiu2020,PhysRevB.103.245435,Li2025SciRep,Nguyen2011JAP}
\begin{align}
P_s&=\frac{G_{\uparrow}-G_{\downarrow}}{G_{\uparrow}+G_{\downarrow}}, \quad
P_v=\frac{G_{K}-G_{K'}}{G_{K}+G_{K'}}.
\end{align}
These quantities provide a direct measure of the imbalance between spin-up and spin-down currents, and between carriers originating from the $\bm K$ and $\bm K'$ valleys, respectively. In the present system, both polarizations are governed by the interplay between the effective Dirac mass $\Delta_{\mathrm{eff}}^{\eta s_z}$ and the channel-dependent longitudinal wave vector $q_x^{\eta s_z}$. The exchange term $m_s$ primarily controls the spin splitting of the bands, thereby enhancing $P_s$, while the staggered potential $V_z$ and the photon-dressed contribution $\lambda_\omega$ introduce a valley-dependent asymmetry that directly impacts $P_v$. 
The barrier width $L$ functions as a geometric tuning parameter that enables selective spin-valley channel enhancement or suppression through Fabry--P\'erot interference. The $P_s$ and $P_v$ values display strong oscillations that depend on $L$ while showing abrupt changes at the boundary between propagating and evanescent wave conditions. 
The same interpretation holds here. Strong irradiation increases the differences between the barrier masses of the four channels. Their propagating windows therefore become increasingly distinct. By tuning $(m_s,V_z)$, one channel can be given a substantially wider propagation window than the others. The barrier width $L$ affects only the accumulated phase. It therefore modifies the resonance sharpness without changing the underlying channel separation.

\begin{figure}[ht!]
\centering 
\includegraphics[scale=0.46]{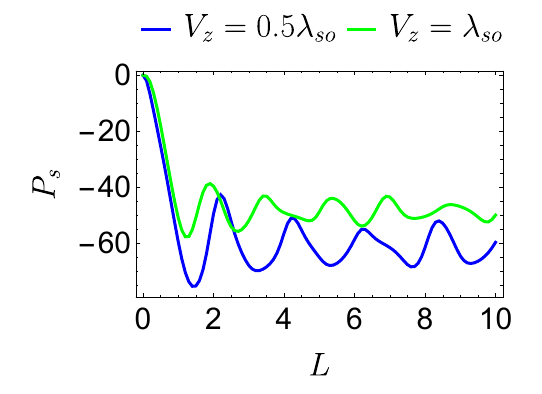}
\includegraphics[scale=0.46]{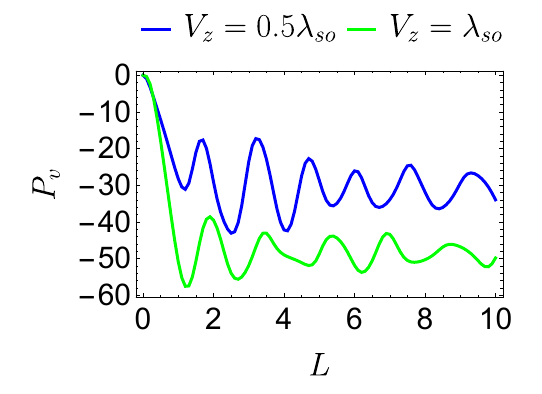}
\caption{Spin and valley  polarization $P_s$, $P_v$ as a function of the barrier width $L$. For $V_0 = 0.5$~eV, $E^{\eta s_z} = 1.1$ eV, $m_s = 0$ eV, $A_0 = 0.8$~Vfs/nm, $V_z = 0.5 \lambda_{\mathrm{so}}$ (blue curve) and $V_z = \lambda_{\mathrm{so}}$ (green curve).}
   \label{fig:PL}
\end{figure}

Figure~\ref{fig:PL} shows the spin and valley polarizations, $P_s$ and $P_v$, as functions of the barrier width $L$. We fix strong driving $A_0=0.8~\mathrm{Vfs/nm}$, and set the exchange to zero, $m_s=0$. Thus far, we have discussed transmission and conductance, whereas here we summarize the same physics at the polarization level, where the $L$ dependence becomes explicit. Only the barrier is photon-dressed; changing $L$ mainly modifies the phase $q_x^{\eta s_z}L$ accumulated in region II, and consequently both $P_s$ and $P_v$ oscillate with $L$. This is the same Fabry--P\'erot physics discussed earlier for $T^{\eta s_z}$ and $G_{\eta s_z}$. There are two trends that stand out. First, both $P_s$ and $P_v$ remain negative over the full range of $L$. This means that the transport is biased toward spin down, $G_{\downarrow}>G_{\uparrow}$, and toward the $K'$ valley, $G_{K'}>G_{K}$. Thus, even without exchange, strong driving produces a systematic preference for a specific spin and valley channel.  
Both polarization signs can be understood from Eq.~\eqref{eq:effgap}. At $m_s=0$, the effective barrier mass is
$\Delta^{\eta s_z}_{\mathrm{eff}}
=\eta(s_z\lambda_{\mathrm{so}}+\lambda_\omega)+V_z.$
For spin-up, the drive and spin--orbit contributions add. For spin-down, they partially cancel. Thus, the spin-down channels have smaller values of $|\Delta^{\eta s_z}_{\mathrm{eff}}|$ in both valleys. Similarly, for $V_z>0$, the positive $\bm K$ masses increase in magnitude, whereas the negative $\bm K'$ masses move closer to zero. The $\bm K'$ channels therefore have smaller mass magnitudes for both spins. Since smaller $|\Delta^{\eta s_z}_{\mathrm{eff}}|$ gives wider propagation windows, the $\bm K'$ and spin-down channels are favored. This explains why both polarizations remain negative.
Second, $V_z$ shifts the balance between spin and valley selectivity. For $V_z=0.5\lambda_{\mathrm{so}}$ (blue), $P_s$ reaches larger negative values and shows stronger oscillations with $L$, while $P_v$ remains more moderate. Increasing $V_z$ to $\lambda_{\mathrm{so}}$ (green) reduces the magnitude of $P_s$, but makes $P_v$ more negative and comparatively steadier. This indicates that $V_z$ can be used to move the operating point between spin-dominated and valley-dominated filtering. Taken together, Fig.~\ref{fig:PL} shows that $L$ acts as a geometric knob that tunes the polarization through interference, while $V_z$ sets which polarization is enhanced under strong photon dressing.

\begin{figure}[ht!]
\centering 
\includegraphics[scale=0.34]{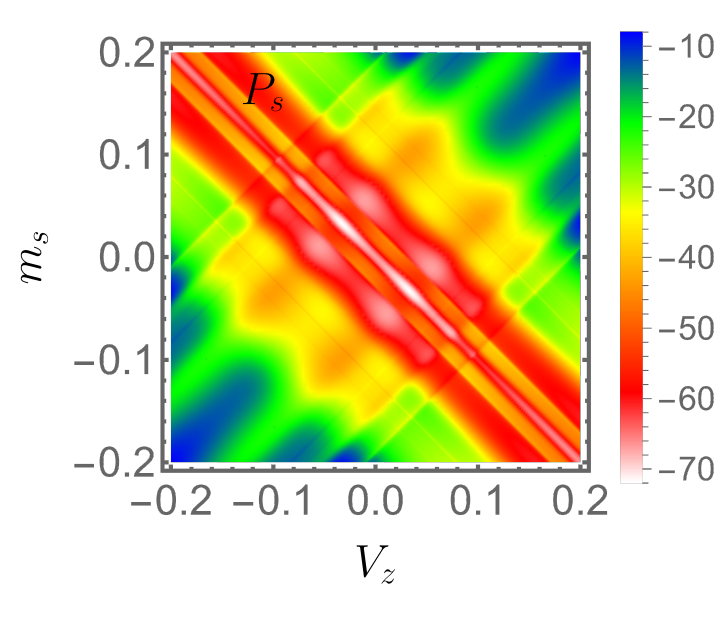}
\includegraphics[scale=0.34]{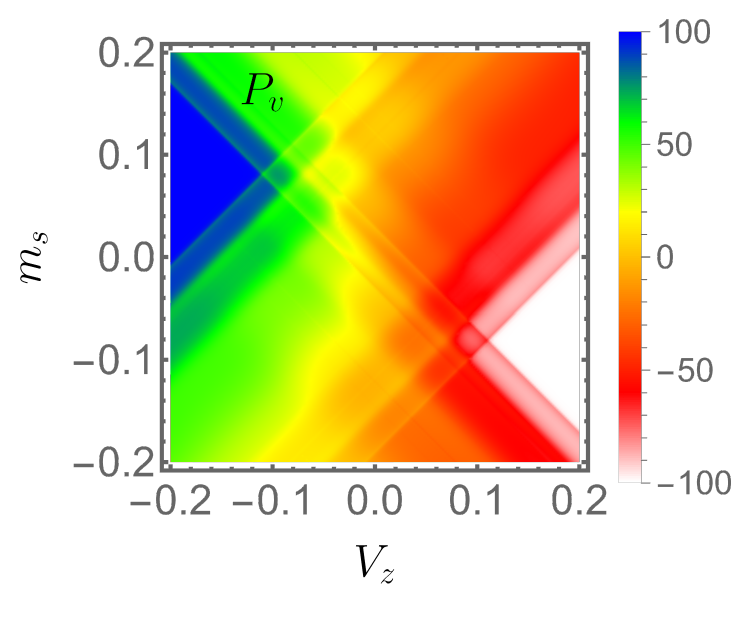}
\caption{Density plot of spin and valley polarization $P_s$ and $P_v$ as functions of the substrate mass $m_s$ and the sublattice staggered potential $V_z$ for $V_0 = 0.5$~eV, $E^{\eta s_z} = 1.1$~eV, $L = 5$ nm, and $A_0 = 0.8$~V$\cdot$fs/nm.}
    \label{fig:PVzms}
\end{figure}

Figure~\ref{fig:PVzms} shows the spin and valley polarizations $P_s$ and $P_v$ in the $(m_s,V_z)$ plane. The color bars report the signed polarizations in percent. This scan, therefore, indicates where the transmitted current is dominated by a particular spin channel and/or a particular valley. We first examine the spin panel. $P_s$ remains negative throughout the plotted range, indicating that $G_{\downarrow}>G_{\uparrow}$ for all considered $(m_s, V_z)$, which confirms what we have seen at the conductance level. The magnitude is not uniform. A diagonal band stands out where the spin imbalance is strongest and $P_s$ reaches values close to $-70\%$, while it is reduced away from this band. This reflects how $(m_s, V_z)$ enters the barrier mass term and reshapes the spin-dependent propagation and matching in the driven region. A weaker modulation is superimposed on this trend, arising from the finite barrier, although it does not change the sign. The spin preference is therefore robust in this regime. The valley panel shows a different pattern. Here, the polarization $P_v$ changes sign across the parameter space. Near-perfect valley filtering occurs where broad parameter regions exhibit $P_v \approx \pm 100\%$, in qualitative agreement with~\cite{Hajati2025}. In these regimes, the current is almost entirely carried by a single valley. By tuning $(m_s,V_z)$ the junction can therefore switch between $\bm K$-dominated transport ($P_v>0$) and $\bm K'$-dominated transport ($P_v<0$). The wedge-shaped boundaries mark the crossover between these regimes. They indicate that the valley contrast can change rapidly under small parameter variations. Taken together, Fig.~\ref{fig:PVzms} highlights operating windows where large spin and valley polarizations occur simultaneously. These regions provide practical working points for spin-valley filtering controlled by $(m_s,V_z)$ under strong photon dressing.

 It is useful to place our results in the context of earlier work on optically dressed Dirac lattices. Under linearly polarized off-resonant fields, the Dirac cones of graphene and the dice lattice acquire elliptical cross sections~\cite{PhysRevResearch.2.043245,PhysRevB.105.115309}.
This shifts the axis of Klein tunneling away from normal incidence and modulates the current transmitted through a barrier. Circularly polarized light instead opens gaps at the Dirac points and transforms the $\alpha$-$T_3$ lattice into a Haldane-like Chern insulator~\cite{PhysRevB.99.205429}. Between the graphene and dice limits, the flat band itself deforms, leaving clear signatures in the optical conductivity~\cite{PhysRevB.107.195137}. However, these effects act identically on the two spin species. A finite spin polarization requires barrier masses that distinguish the two spins, which none of these models provides.

Silicene junctions do provide such masses, since the coupling there has the Kane--Mele form, and an irradiated junction between undriven leads indeed shows spin- and valley-polarized conductance~\cite{Zhang2019Silicene}. That result was obtained in the terahertz regime, where the photon energy is comparable to the gaps and the polarization arises from selective suppression at sideband anticrossings, which are pinned to multiples of half the photon energy. Here the drive is off-resonant, so the sidebands remain unpopulated and the light acts only through the static mass. The intrinsic spin--orbit coupling of jacutingaite, $\lambda_{\mathrm{so}}\simeq 81$~meV, is more than twenty times the silicene value of $3.9$~meV, so the barrier masses of the two spins differ by $2\lambda_{\mathrm{so}}\simeq160$~meV. The polarizations shown in Figs.~\ref{fig:PL} and \ref{fig:PVzms} then extend over broad parameter windows.

\section{Conclusion}
\label{sec:conclusion}

We studied transport through a monolayer jacutingaite tunnel junction in which only the barrier is irradiated by off-resonant circularly polarized light. In the high-frequency regime, the driven barrier reduces to an effective static problem with a valley-dependent photon-dressed mass term. The four spin-valley channels then acquire different effective gaps in the barrier and therefore different propagation conditions and scattering phases.

Two ingredients generate selectivity: the first is photon dressing. It shifts the effective barrier {Dirac mass} with opposite sign in $\bm K$ and $\bm K'$, {thereby driving} the propagation threshold {in} opposite directions in the two valleys. This creates windows where one valley is propagating while the other is evanescent. The second ingredient is the finite barrier width. Once a channel is propagating, transport is controlled by the phase accumulated across the barrier, $q_x^{\eta s_z}L$. Since $q_x^{\eta s_z}$ depends on the spin-valley channel, the resonance condition is channel dependent. The barrier, therefore, acts as a Fabry--P\'erot cavity in a channel-dependent way. In practice, $L$ becomes a geometric knob: one can place a target channel on resonance while pushing competing channels off resonance.

The above effects survive angular averaging, i.e., after the $k_y$ integration in the conductance. The Landauer conductance retains a clear valley and spin contrast. The polarization measures show extended parameter ranges with large $P_v$ and sizable $P_s$. The $(m_s, V_z)$ maps summarize the main knobs. Photon dressing primarily sets the valley contrast, while exchange enhances the spin contrast. When both are present, one finds regimes where both polarizations are simultaneously large. In this sense, the junction acts as a simple opto-spintronic element, where light and substrate engineering steer the dominant spin-valley channel.

Our analysis focuses on the off-resonant, prethermal regime, where a leading-order high-frequency description is appropriate. Natural extensions include finite temperature, disorder, and interface roughness. Another direction is the intermediate-frequency regime, where Floquet sidebands must be included explicitly in a full Floquet-scattering treatment. Multi-terminal geometries and realistic contact modeling would further connect the predicted filtering windows to device-level spin-valley valve operation and experimental readout.

\section*{Data Availability Statement}
The data that support the findings of this study are available from the corresponding author upon reasonable request.

\section*{Acknowledgment}
 O. Bouladiane and K. Azaidaoui acknowledge the support provided by CNRST in the framework of the program "PhD-Associate Scholarship –- PASS".

\bibliographystyle{apsrev4-2}
\bibliography{literature}

\end{document}